\documentclass[11pt]{article}
\usepackage[margin=1in]{geometry}
\usepackage{graphicx,amsmath,amssymb,amsfonts,dsfont,physics,authblk}
\usepackage{multirow}
\usepackage{tikz}
\usepackage{verbatim,braket}
\usepackage{empheq}
\usepackage[sort&compress, numbers]{natbib}
\newcommand{\eprint}[2][]{\href{https://arxiv.org/abs/#2}{arXiv:~\nolinkurl{#2}}}

\usepackage[hidelinks]{hyperref}
\usepackage{xcolor}
\usetikzlibrary{decorations.pathreplacing, decorations.pathmorphing, patterns, hobby, decorations.markings, positioning, matrix,shapes,arrows,fit,automata}
\usepackage{booktabs}
\usepackage{pgfplots}
\pgfplotsset{compat=1.3}
\pgfkeys{/pgfplots/tuftelike/.style={
  semithick,
  tick style={major tick length=4pt,semithick,black},
  separate axis lines,
  axis x line*=bottom,
  axis x line shift=10pt,
  xlabel shift=-5pt,
  axis y line*=left,
  axis y line shift=10pt,
  ylabel shift=-5pt}}

\tikzset{
    scattering/.style={
        baseline=-0.5ex,
        x=0.8cm,y=0.8cm,
        line width=0.5pt,
        every node/.style={font=\scriptsize,inner sep=1pt},
        decoration={snake,amplitude=0.45mm,segment length=1.6mm}
    }
}

\usepgfplotslibrary{colorbrewer, groupplots}

\pgfplotsset{
    colormap/inferno/.style={%
    /pgfplots/colormap={inferno}{%
      rgb=(0.001462, 0.000466, 0.013866)
      rgb=(0.037668, 0.025921, 0.132232)
      rgb=(0.116656, 0.047574, 0.272321)
      rgb=(0.217949, 0.036615, 0.383522)
      rgb=(0.316282, 0.053490, 0.425116)
      rgb=(0.410113, 0.087896, 0.433098)
      rgb=(0.503493, 0.121575, 0.423356)
      rgb=(0.596940, 0.154848, 0.398125)
      rgb=(0.688653, 0.192239, 0.357603)
      rgb=(0.775059, 0.239667, 0.303526)
      rgb=(0.851384, 0.302260, 0.239636)
      rgb=(0.912966, 0.381636, 0.169755)
      rgb=(0.956852, 0.475356, 0.094695)
      rgb=(0.981895, 0.579392, 0.026250)
      rgb=(0.987464, 0.690366, 0.079990)
      rgb=(0.973088, 0.805409, 0.216877)
      rgb=(0.947594, 0.917399, 0.410665)
      rgb=(0.988362, 0.998364, 0.644924)
    },
  },
}

\definecolor{lightred}{RGB}{251,154,153}
\definecolor{darkblue}{RGB}{31,120,180}
\definecolor{lightgreen}{RGB}{178,223,138}
\definecolor{darkred}{RGB}{227,26,28}
\definecolor{darkgreen}{RGB}{51,160,44}
\definecolor{lightblue}{RGB}{166,206,227}
\definecolor{lila}{RGB}{106,61,154}
\definecolor{myorange}{RGB}{255,127,0}

\newcommand{\spin}{\mathbf{S}}

\DeclareMathOperator\arccosh{arccosh}

\begin{document}

\title{Field theory and the biquadratic spin-1 chain }

\author[1,2]{Luisa Eck}
\author[2,3]{Paul Fendley}
\affil[1]{\small Walter Burke Institute for Theoretical Physics,
California Institute of Technology, Pasadena, California 91125, USA}
\affil[2]{\small Rudolf Peierls Centre for Theoretical Physics, Parks Rd, Oxford OX1 3PU, United Kingdom}
\affil[3]{\small All Souls College, Oxford, OX1 4AL, United Kingdom}

\date{\today} 

\maketitle

\begin{abstract}
The antiferromagnetic biquadratic spin-1 chain is gapped, $SU(3)$ symmetric, and integrable. Standard arguments suggest that its low-energy physics is described by the $\mathbb{CP}^2$ nonlinear sigma model at $\theta$\;=\;$\pi$. 
Using the Bethe ansatz, we show that its low-energy excitations can be interpreted as deconfined domain walls,
just as in the sigma model. 
However, the lattice model is integrable, whereas the field theory is generally believed not to be. 
We propose a solution to this mystery by finding an integrable field theory with the same low-energy limit but which is not Lorentz invariant.
The integrability-preserving scaling limit is obtained by adding a higher conserved charge and tuning the chain to criticality. The gap closes quadratically, so that the limit must be described by some nonrelativistic field theory rather than the relativistic sigma model. We describe the two-particle scattering matrix of this field theory. 
\end{abstract}

\section{Introduction}

Two-dimensional $\mathbb{CP}^{m-1}$ nonlinear sigma models have long been studied as toy models for four-dimensional gauge theories. They exhibit many of the same phenomena, including asymptotic freedom, instantons, and dependence on a topological $\theta$ angle \cite{DAdda1978vbw, WITTEN1979285}. At $\theta=\pi$ and for $m > 2$, there are two degenerate vacua that spontaneously break charge conjugation symmetry, similar as in 4d $SU(m)$ Yang-Mills theory at $\theta=\pi$. The domain wall excitations are then deconfined \cite{seiberg1984topology}, just as in the massive Schwinger model \cite{coleman1976schwinger}.
At generic $\theta$, the excitations are confined because they are separated by a region of higher energy.
From a modern perspective, the sigma model at $\theta=\pi$ exhibits a mixed anomaly between charge conjugation and $PSU(m)$ symmetry for even $m$ and a global inconsistency for odd $m$ \cite{gaiotto2017theta,ohmori2019sigma}. 

A profound connection to condensed-matter physics came from the recognition that these field theories give continuum descriptions of antiferromagnetic quantum spin chains. Haldane famously showed that the low-energy physics of the $SU(2)$ Heisenberg chain is described by the $\mathbb{CP}^1$ (also named $O(3)$, for its global symmetry group) sigma model at $\theta=0$ for integer spin and $\theta=\pi$ for half-integer spin \cite{HALDANE1983464,Affleck1987}. Affleck extended this correspondence to a family of $SU(m)$ symmetric spin chains \cite{affleck1985large}. 
This semiclassical mapping is controlled for large onsite representations, with the number $n_c$ of columns in the Young diagram going to infinity at fixed $m$ \cite{Affleck:1984ar}.
Later work gave a spin coherent-state formulation of this mapping \cite{affleck2022flag}.
Read and Sachdev developed a complementary large-$m$ approach using bosonic coherent states for the Schwinger bosons, taking $m$ and $n_c$ both to infinity with $n_c/m$ fixed
\cite{readsachdev1989,Read1990}.
Separately, large-$m$ calculations for the chain with alternating
fundamental and antifundamental representations, which
are the smallest nontrivial representations of $SU(m)$, support a
qualitative correspondence with the $\mathbb{CP}^{m-1}$ model
at $\theta=\pi$ \cite{affleck1985large}.

In this paper, we focus on a particularly intriguing case, the  $SU(3)$-invariant ``biquadratic" spin-1 chain 
\cite{parkinson1987integrability, parkinson1988biquadratic, barber1989biquadratic, Kluemper_1989}.
While a fair amount of work has been done on this chain \cite{sorensen1990, affleck1990exact, Xian1993dimer, read2007enlarged, Aufgebauer2010QuantumSC}, 
a central mystery remains. Standard arguments suggest that its low-energy theory should be the same as that of the $\mathbb{CP}^2$ field theory with $\theta=\pi$ \cite{Affleck:1984ar}.
However, the biquadratic chain is integrable, while this field theory is believed not to be. While there is no conclusive proof of the latter statement, early papers found that classical nonlocal conserved charges cannot be promoted to quantum charges \cite{abdalla1981anomaly, Gomes:1982qh}.
Moreover, attempts to construct an exact scattering matrix for this field theory fail. Namely, integrable field theories such as the $O(m)$ sigma model are characterized by a two-particle $S$-matrix satisfying strong constraints 
\cite{Zamolodchikov1978factorized}.
However, a putative $S$ matrix for $\mathbb{CP}^{m-1}$ at $\theta=\pi$ encounters obstructions once relativistic crossing symmetry is imposed \cite{Fendley:2002ww}.  

Of course it is unsurprising if a field theory is integrable but has a lattice analog that is not, as it is possible that only integrable degrees of freedom survive in the scaling limit. The reverse, however, is not possible: if the $\mathbb{CP}^{2}$ model at $\theta$\,=\,$\pi$ is  not integrable, then it cannot be an effective field-theory description of the spin chain, as any sensible limit of the latter must remain integrable. 
Nevertheless, a considerable amount of work indicates that the low-energy physics at the very least has the same qualitative properties \cite{affleck1985large, read1989phasediagram, beard2005cluster}.
Here we confirm and extend these results.

Another difficulty in finding a field-theory limit of the biquadratic chain is that it is gapped \cite{aizenman2020dimerization}, 
whereas e.g.\ the spin-$\tfrac12$ Heisenberg model and the corresponding $\mathbb{CP}^1$ field theory with $\theta=\pi$ are gapless. Thus in identifying the biquadratic chain with a field theory one cannot use the powerful tools of conformal invariance in any apparent way. As far as we are aware, all known integrable sigma models with $\theta$\,=\,$\pi$ are gapless; see e.g.\ \cite{Fendley2000}. There are thus no analogous situations from which we can gain guidance. Making the situation even subtler is that a priori it is not obvious how to deform the model to the region of a critical point so as to make a field-theory description possible. The gap increases with $m$, so taking the large-$m$ limit does not help.

We here propose a resolution of these mysteries. Our first result is to show that there is no easy way out: we complete the identification of the qualitative properties of the quasiparticles of the biquadratic chain with those of the sigma model. 
We show in particular that in addition to its long-known properties of spontaneous dimerization and a gap, this chain exhibits {\em deconfinement}. We then find an integrability-preserving deformation to a critical point that should admit a field-theory description. However, this point is not Lorentz-invariant and the sigma model cannot apply.

The correspondence between the qualitative low-energy properties does remain striking, but demonstrating so requires some effort.
The two dimerized ground states of the biquadratic chain are degenerate only in the thermodynamic limit, and finding 
properties of the low-energy excitations is even trickier because of the large correlation length $\xi \approx 42$. Given the two ground states, it is natural to expect that the lowest-energy excitations are kinks separating them. As we review in Section \ref{subsec:domainwalls}, each such kink forms a threefold multiplet, alternating between 3 and $\overline{3}$ representations of the $SU(3)$ symmetry. A two-particle state then will transform as $3\otimes \overline{3}=8+1$ under $SU(3)$. The key physics question then is whether these excitations are confined or deconfined, i.e.\ whether or not the energy grows as this pair is separated. If deconfined, all nine two-particle states must be degenerate, up to finite-size corrections. (In an integrable model, the eigenstates are in the form of plane waves, and so the energy takes on the value of widely separated states.)  If confined, the pair cannot be separated, and so the octet and singlet states will differ in energy. 

We indeed establish that this degeneracy occurs in the thermodynamic limit of the biquadratic chain, utilizing the Bethe ansatz and a sector-wise mapping to the antiferromagnetic XXZ chain. Thus all seems to be in harmony with the field theory: in the $\mathbb{CP}^{m-1}$ sigma models, the expected behaviour (established at large $m$ and for $m$\,=\,2) is that excitations are confined when  $\theta$\,=\,0, while they are deconfined when $\theta$\,=\,$\pi$. Nonetheless, this calcluation does not resolve the mystery of the integability.  

To probe further, we study a continuum limit that preserves integrability. Rather than taking the large-representation limit, we instead 
perturb the biquadratic chain by including a higher conserved charge in the Hamiltonian. Using the mapping to the XXZ chain, we find this integrable deformation has both a massless limit around which one can build a field theory and, crucially, the same low-energy quasiparticles as the $\mathbb{CP}^{2}$ model at $\theta$\,=\,$\pi$. However, this field theory is not Lorentz invariant like the sigma model because the dispersion at the critical point is quadratic. Its scattering matrix avoids the inconsistency in the earlier attempt because crossing symmetry only occurs in Lorentz-invariant field theories.

The remainder of the paper is organized as follows.
Section~\ref{sec:biquadratic} introduces the chains of interest, and gives their precise connection to the XXZ chain using the affine Temperley-Lieb algebra. We exploit these results in Section~\ref{sec:bethe} to derive the low-energy properties of the biquadratic chain, including deconfinement. 
Section~\ref{sec:fieldtheory} reviews the $\mathbb{CP}^{m-1}$ nonlinear sigma models at $\theta=\pi$, including the evidence against quantum integrability. 
Section~\ref{sec:perturbation} introduces the higher-charge perturbation and analyzes its critical point described by a nonrelativistic field theory. Symmetry and integrability constrain its S-matrix, which we expect to be closely related to that of the massive XXZ chain.

\section{The biquadratic spin-1 chain and its XXZ avatar}
\label{sec:biquadratic}

The biquadratic chain attracted early interest as an integrable and near-critical point in the bilinear--biquadratic spin-1 phase diagram \cite{solyom1987biquadratic, singh1988spin1, Affleck:1985exact}. 
It has an $SU(3)$ symmetry, with the spins alternating between fundamental and antifundamental representations. The most general nearest-neighbor interaction therefore can be written as a projector onto the singlet state. Such projectors generate the Temperley-Lieb algebra \cite{Temperley1971}. Since the spin-$\tfrac12$ XXZ chain provides another presentation of the same algebra, the spectra of the two chains can be related sector by sector \cite{Aufgebauer2010QuantumSC}. 

The mapping between the biquadratic chain and the XXZ chain must be treated with some care. The Hilbert space of the former is exponentially larger: for system length $L$ it grows as $3^L$, while that of the XXZ chain grows only as $2^L$. Exponentially large degeneracies in the spectrum of the former must occur even at finite sizes. Read and Saleur provided valuable insight into these degeneracies by showing that the biquadratic chain and their $m$-state generalizations have a much larger symmetry than $SU(m)$ \cite{read2007enlarged}. Intriguingly, this symmetry persists even for disordered couplings, where integrability is lost. In this section we give these degeneracies, making contact between the results of \cite{read2007enlarged} and \cite{Aufgebauer2010QuantumSC}. We exploit this correspondence between chains in Section~\ref{sec:bethe} to compute the low-energy spectrum.

\subsection{\texorpdfstring{The biquadratic chain and its $SU(m)$-symmetric generalizations}
{The biquadratic chain and its SU(m)-symmetric generalizations}}

The biquadratic spin-1 chain has built-in $SU(2)$ symmetry, enhanced to $SU(3)$ by the particular choice of Hamiltonian
\begin{equation}\label{eq:Hbq}
H_\mathrm{bq}
= -\sum_{j=1}^L (\mathbf S_j\!\cdot\!\mathbf S_{j+1})^2
= -\sum_{j=1}^L \bigl(3P_j^{(0)} + 1\bigr),
\end{equation}
where the $\mathbf{S}_j$ are the usual $SU(2)$ spin-1 generators acting non-trivially on the $j$th site, and
\begin{equation}\label{eq:Pbq}
P_j^{(0)}=\frac13\left[(\mathbf S_j\!\cdot\!\mathbf S_{j+1})^2-1\right]
\end{equation}
projects two adjacent spin-1's onto their total spin-0 subspace. We take an even number of sites $L$, along with periodic boundary conditions. 

The $SU(3)$ symmetry becomes manifest by rewriting the biquadratic spin-1 chain as the $m=3$ member of a family of $m$-state spin chains with $SU(m)$ symmetry.
Even sites transform in the fundamental representation $m$, while odd sites transform in the antifundamental
representation $\overline{m}$. The only nearest-neighbor Hamiltonian invariant under $SU(m)$ is the sum over projectors onto the singlet for each pair. It can be written in terms of the symmetry generators as
\cite{affleck1990exact,read2007enlarged}
\begin{equation}\label{eq:hamSUm}
H_m =
\frac12\sum_{i=1}^{L/2}\sum_{a=1}^{m^2-1}
\left(
\bar\lambda_{2i-1}^a\lambda_{2i}^a
+\lambda_{2i}^a\bar\lambda_{2i+1}^a
\right)-\frac{L}{m},
\end{equation}
where $\lambda^a$, $a=1,\dots,m^2-1$, are now $SU(m)$ generators in the fundamental representation, normalized by $\operatorname{tr}(\lambda^a\lambda^b)=2\delta^{ab}$,
and $\bar\lambda^a=-(\lambda^a)^*$ are the corresponding generators in the antifundamental representation. For $m=3$, the $\lambda^a$ may be chosen as the Gell-Mann matrices, and $H_{\mathrm{bq}}=H_3-L$.

Worth noting is the other $SU(3)$-symmetric integrable spin-1 chain in the bilinear-biquadratic phase diagram \cite{sutherland1975spinchain, itoi1997spin1}. It has the fundamental representation on every site and is gapless. Its physical properties are thus very different.

\subsection{Dimer ground states and deconfined domain wall excitations}\label{subsec:domainwalls}

\begin{figure}
    \centering
    \includegraphics{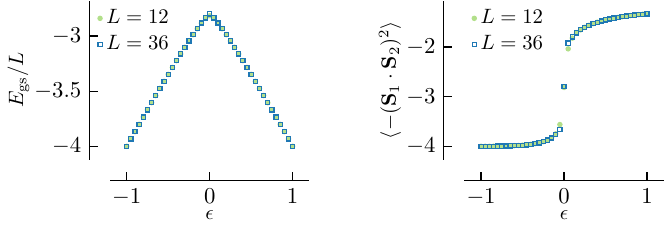}
    \caption{\small The first-order transition at the unstaggered point $\epsilon = 0$ is apparent in the ground-state energy (left) and the bond energy (right). Finite-size effects are small for periodic boundary conditions used here.}
    \label{fig:firstorder}
\end{figure}

The biquadratic chain is gapped, as first established in the physics literature through the mapping to the XXZ chain described below  \cite{affleck1990exact, barber1989biquadratic, sorensen1990, Xian1993dimer}, and later proven in \cite{aizenman2020dimerization, bjornberg2021dimerization}. The ground states are {\em dimerized}, as illuminated by studying the chain with staggered couplings, namely
\begin{equation}\label{eq:Hstaggered}
    H_\text{bq}(\epsilon)
    =
    -(1+\epsilon)\sum_{j\ \mathrm{odd}}
    \left(\spin_j\cdot\spin_{j+1}\right)^2
    -(1-\epsilon)\sum_{j\ \mathrm{even}}
    \left(\spin_j\cdot\spin_{j+1}\right)^2 \ .
\end{equation}
For $|\epsilon|$ large, the ground state is dominated by states with singlet bonds between sites $j$ and $j+1$ with $j$ odd for $\epsilon$ positive and even for negative. 
One of these two ``dimer" states persists as the unique ground state for any non-zero $\epsilon$ \cite{pruisken2022super}, as can be verified with DMRG; see Fig.~\ref{fig:firstorder}.
At $\epsilon=0$ with periodic boundary conditions, full translation invariance is restored, and the two ground states with momenta 0 and $\pi$ become degenerate. The gapped biquadratic chain thus describes a first-order transition between the two dimerized phases for $\epsilon>0$ and $\epsilon<0$.

Intuition into the lowest excited states follows naturally from this dimer description of the ground states. Namely, these states should be gapped domain walls separating regions of the two ground states. Domain walls on the even sites (kinks) have the $\epsilon=0^-$ ground state to the left of them and $\epsilon=0^+$ to the right, while those on odd sites (antikinks) the reverse. Kinks and antikinks must therefore alternate in space, and so with periodic boundary conditions and even $L$ they must occur in equal numbers.

The $SU(3)$ symmetry provides further illumination. As apparent from writing $H_{\rm bq}$ in the form Eq.~\eqref{eq:hamSUm}, the kinks transform as the fundamental representation $3$ of $SU(3)$, while the antikinks transform  as the antifundamental representation $\bar{3}$.  Since these representations obey $3 \otimes \overline 3=1 \oplus 8$, each kink-antikink pair can be decomposed into singlet and octet states (except when they occupy adjacent sites, as the singlet then is simply the ground state).

Consider an excited state consisting of a widely separated kink-antikink pair, a natural notion given that the system is gapped. A basic physical question is then whether the pair is confined: does its energy grow as the separation increases? In this integrable system, this property follows from the energy difference between the octet and singlet states. If it goes to zero in the thermodynamic limit, the kinks and antikinks are {\em deconfined}. 
When $\epsilon \ne 0$, separating two domain walls creates a region of the unfavored ground state, which can be interpreted as a confining potential.
Consequently, here the singlet and octet states are no longer degenerate even as $L\to\infty$. 

\begin{figure}
    \centering
    \includegraphics{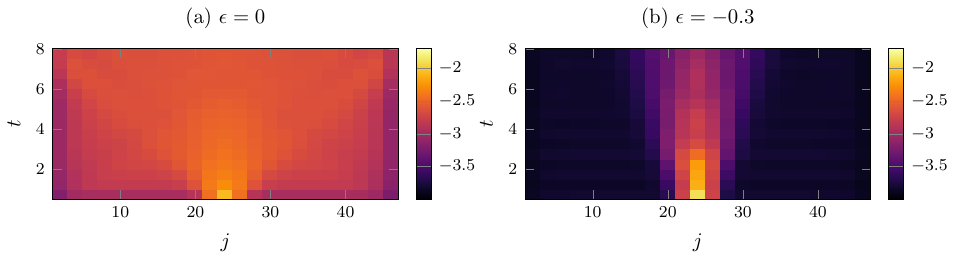}
    \caption{\small Bond energies
    $\langle -(\spin_j\cdot\spin_{j+1})^2\rangle(t)$ after time evolution with the
    staggered Hamiltonian \eqref{eq:Hstaggered}, starting from a dimer state with one
    broken bond in the middle, for $\epsilon=0$ and $\epsilon=-0.3$ ($L = 48$, periodic boundary conditions, maximum bond dimension $2000$).}
    \label{fig:spectrumbiquadratic}
\end{figure}

\begin{figure}
    \centering
    \includegraphics{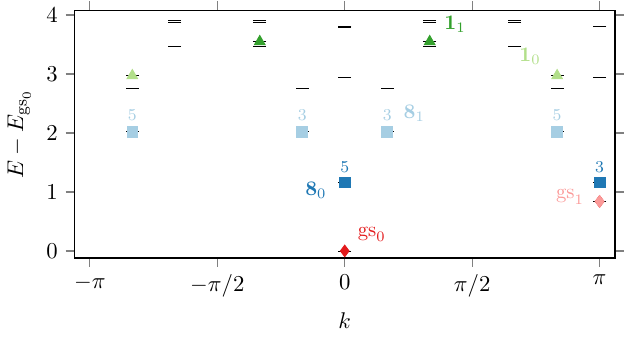}
    \caption{\small Low-energy spectrum of the periodic $L=12$ biquadratic chain,
    from exact diagonalization in the $S^z=0$ sector. The two finite-size
    ground states at $k=0,\pi$, the two lowest octets at $k=0$ and $k=\pi \pm 2\pi/L$, and the two lowest singlets at $k=\pi\pm2\pi/L, \pm 4 \pi/L$ are highlighted. 
    The labels 5 and 3 give the full spin-multiplet degeneracies; a quintuplet and a triplet at momenta differing by $\pi$ together form one $SU(3)$ octet.}
\label{fig:spectrumbiquadratic_new}
\end{figure}

For the unstaggered biquadratic chain, the answer is far from obvious: the connection to the sigma model hints that it should be deconfined, and a qualitative difference between the behaviors for $\epsilon=0$ and $\ne 0$ is apparent in Fig.~\ref{fig:spectrumbiquadratic}.
At $\epsilon=0$, the energy initially concentrated at the broken dimer spreads across the chain, consistent with deconfined domain walls moving apart. With staggering, it remains localized because separating the walls creates a growing region of the energetically disfavored dimerization.
These results were obtained using time-evolving block decimation
(TEBD) implemented with ITensor \cite{itensor}, following the
method of Ref.~\cite{kormos2017real}.

Exact diagonalization is not very illuminating because of the large correlation length. From Fig.~\ref{fig:spectrumbiquadratic_new}, it is apparent that at $L$\,=\,12 the singlet states have much higher energy than not only the lowest-lying octet state, but other states in the spectrum as well. Worth noting as well is that since the representations alternate between 3 and $\overline{3}$, some $SU(3)$ generators have momentum $\pi$.

Nonetheless, we establish in Section~\ref{sec:bethe} that indeed the model is deconfined: the singlet and octet excitations become degenerate in the thermodynamic limit. 
We also show why the exact diagonalization results are not illuminating, because 
the energy splitting between the singlet and octet excitations vanishes approximately as $\sim 1/L^3$.
Before doing these derivations, however, we must first describe the Temperley-Lieb mapping to the XXZ chain, the subject of the next subsection.

\subsection{Affine Temperley-Lieb algebra and mapping to the XXZ chain}\label{subsec:mapping}

A classic result of statistical mechanics is Temperley and Lieb's demonstration of the close interconnections between a number of celebrated models. They showed how a variety of two-dimensional classical transfer matrices and their quantum-chain limits can be written in terms of a set of generators that obey an algebra that now bears their names \cite{Temperley1971}. The Temperley-Lieb algebra connects seemingly very different models, including the XXZ chain, the Potts model and loop models. More generally, one can define such a model for any fusion category \cite{Blakeney2025}. Most importantly for us, the biquadratic chain and its $SU(m)$ generalization can be written in this fashion  \cite{barber1989biquadratic,Aufgebauer2010QuantumSC,read2007enlarged}. 
The Temperley-Lieb structure fixes allowed values of the energy levels, but not their degeneracies. Here we explain the correspondence between the two models and their Hilbert spaces, and in the next section use the Bethe ansatz to compute their energies.

The standard convention for the nearest-neighbour projector on the singlet is to define $e_i=mP_i^{(0)}$, 
so that the $m=3$ biquadratic chain Hamiltonian \eqref{eq:Hbq} is 
\begin{equation}\label{eq:HbqHTL}
    H_{\mathrm{bq}}
    =-\sum_i(e_i+1)\ .
\end{equation}
For the chain with open boundary conditions, the generators $e_i$, $i=1,\dots,L-1$, satisfy the defining relations of the Temperley-Lieb algebra with parameter $m$ \cite{Temperley1971}:
\begin{equation}\label{eq:tl}
e_i^2 = m e_i, \quad e_i e_{i \pm 1} e_i = e_i, \quad \big[e_i, e_j\big] = 0 \quad \text{for } |i - j| > 1.
\end{equation}

A key observation is that Temperley-Lieb Hamiltonians can have different Hilbert-space dimensions. Since the spectra of different models match sector by sector, the model with the larger dimension will typically have exponentially large degeneracies.
\citet{read2007enlarged} showed that the Temperley-Lieb spin chains we study possess a symmetry algebra much larger than $SU(m)$ for $m\ge 3$, accounting presumably for all these degeneracies. This enlarged symmetry is the commutant $\mathcal A$ of the Temperley-Lieb algebra $\mathcal C$, namely the algebra of operators that commute with all local generators $e_i$. The Hilbert space consequently decomposes into representations of $\mathcal A \times \mathcal{C}$. Here we use this decomposition to relate each Temperley-Lieb sector to the corresponding sector of the spin-$\tfrac12$ XXZ
chain, following \citet{Aufgebauer2010QuantumSC}. 

Writing $q+q^{-1}=m=2\Delta$, the XXZ presentation of
the Temperley--Lieb generators is
\begin{equation}\label{eq:ei_XXZ}
\begin{aligned}
e_i^{\mathrm{XXZ}}
={}-\left(
e^{i\frac{\phi}{L}}\sigma_i^+\sigma_{i+1}^-
+e^{-i\frac{\phi}{L}}\sigma_i^-\sigma_{i+1}^+
\right)
-\frac{\Delta}{2}
\left(\sigma_i^z\sigma_{i+1}^z-1\right)
+\frac{q-q^{-1}}{4}
\left(\sigma_{i+1}^z-\sigma_i^z\right).
\end{aligned}
\end{equation}
Here $\phi$ is a total boundary twist, distributed uniformly over the chain.
The last term telescopes to zero for periodic boundary conditions, giving
\begin{equation}\label{eq:hamXXZ}
H_{\mathrm{XXZ}}(J,\Delta,\phi)
=-\frac{J}{2}\sum_i e_i^{\mathrm{XXZ}}.
\end{equation}
Thus $J=2$ realizes the algebraic Hamiltonian $-\sum_i e_i$.
The biquadratic Hamiltonian in \eqref{eq:Hbq} contains an
additional constant $-L$, so corresponding eigenvalues satisfy
$E_{\mathrm{bq}}=E_{\mathrm{XXZ}}-L$.

For the Bethe ansatz analysis below in Section~\ref{sec:bethe}, periodic boundary conditions are more convenient than open ones. The relevant algebra is then the affine Temperley--Lieb algebra, obtained by extending \eqref{eq:tl} with the generator $e_L$
and a one-site translation generator $u$ \cite{levy1991periodic,pasquier1990quantumgroups}:
\begin{gather}\label{eq:jtl}
    e_i^2 = m e_i, \qquad
    e_i e_{i\pm1} e_i = e_i, \qquad
    [e_i,e_j] = 0
    \quad \text{for } 1<|i-j|<L-1, \\
    u e_i u^{-1} = e_{i+1}, \qquad
    u^2 e_{L-1} = e_1 e_2 \cdots e_{L-1}.
\end{gather}
Here $i,j=1,\ldots,L$, with generator indices understood modulo $L$, and $L$ is assumed to be even throughout.
For the alternating $SU(m)$ chain, we use the subalgebra generated
by the $e_i$ and $u^{\pm2}$, since two-site translation preserves
the fundamental and antifundamental sublattices.
Unlike the open Temperley-Lieb algebra, the affine Temperley-Lieb algebra is infinite-dimensional. This property is readily apparent in the graphical presentation, where an arbitrary number of lines can wind around the torus.

The affine Temperley-Lieb algebra has finite-dimensional standard modules $W_{j,z^2}$, which are labeled by the number $2j$ of through-lines in the diagrammatic presentation and a complex twist parameter $z\in\mathbb C$. 
In the $SU(m)$ spin chains, each through-line represents a spin that is not paired with another spin into an $SU(m)$ singlet.
For $j > 0$, the twist parameter $z$ records the phase picked up when through-lines wind around the periodic direction
\citep{Martin:1993jka,Gainutdinov:2017aom}. 
This corresponds to cyclically permuting the unpaired spins while leaving the singlet
pairings unchanged \citep{read2007enlarged}.
The dimensions of these modules depend only on $L$ and $j$, not on $m$ and $z$: 
\begin{equation}\label{eq:dhat}
    \hat d_j=\binom{L}{L/2+j}, \qquad j=0,1,\dots,L/2.
\end{equation}
The Hilbert space of the $SU(m)$ chain then can be decomposed as
\citep{read2007enlarged,moudgalya2022frag}
\begin{equation}\label{eq:HilbTL}
    \mathcal{H}_{SU(m)}
    =
    \bigoplus_{j=0}^{L/2}
    \left(
    V_{j}^{(\mathcal A)}
    \otimes
    W_{j,z^2}^{(\mathcal C)}
    \right),
\end{equation}
where $V_j^{(\mathcal A)}$ is a $\hat{D}_j$-dimensional irreducible representation of the commutant. The commutant algebra has representations with dimensions \citep{Aufgebauer2010QuantumSC}
\begin{equation}\label{eq:Dj}
    \hat{D}_j = \begin{cases} 1 & j=0 \\  q^{2j} + q^{-2j} & j \geq 1 \end{cases} \quad \text{ with } q+q^{-1} = m,
\end{equation}
so in particular $\hat{D}_1 = m^2 - 2$.

The Hilbert-space decomposition used by \citet{read2007enlarged} is slightly different. 
They take the Jones--Temperley--Lieb quotient of the affine Temperley-Lieb algebra, in which $u^L=1$.
Diagrammatically, this identifies a full
collective winding of the through-lines with no winding, requiring $z^{2j}=1$ for $j>0$.
For $j=0$, there are no through-lines, so closed loops can wind around the periodic direction. Matching their weight to that of the
$SU(m)$ spin chain requires $z+z^{-1}=m$.

For $j>0$, passing to the Jones--Temperley--Lieb quotient leaves the module dimensions \eqref{eq:dhat} unchanged.
In the $j=0$ sector, choosing $z=q$ gives the affine standard module $W_{0,q^2}$.
Its Jones--Temperley--Lieb quotient $\overline W_{0,q^2}$ has the smaller dimension
\begin{equation}\label{eq:jtld0}
    \overline{d}_0 = \binom{L}{L/2} - \binom{L}{L/2 + 1}.
\end{equation}
This is compensated by a larger dimension of the $j=1$ representation of the commutant, 
\begin{equation*}
    \overline{D}_1 = q^2 + q^{-2} + 1 = m^2 -1.
\end{equation*}
Now $\overline{D}_1$ matches the dimension of the adjoint representation of $SU(m)$, and the commutant enlarges the $SU(m)$ symmetry.
Although the affine Temperley–Lieb and Jones–Temperley–Lieb decompositions organize the sectors of the Hilbert space differently, they both reproduce the full Hilbert-space dimension $m^L$. In particular,  with $\overline{D}_0=\hat{D}_0=1$,
\begin{equation*}
    \hat{d}_0 \hat{D}_0 + \hat{d}_1 \hat{D}_1 = \overline d_0 \overline D_0 + \overline d_1 \overline D_1.
\end{equation*}
For open chains, the corresponding dimensions of the commutant representations were recently rederived using a recursion for the Jones-Wenzl projectors \cite{ferydouni2026spectrum}.

It remains to identify the twists in \eqref{eq:ei_XXZ} that reproduce the relevant sectors of the biquadratic chain.
The XXZ chain has a $U(1)$ symmetry generated by $S^z = \frac{1}{2} \sum_i \sigma^z_i$, which commutes with each $e_i$ and with the square of the translation operator $u^2$. Its Hilbert space decomposes into magnetization sectors as
\begin{equation}\label{eq:Hilbxxz}
\mathcal{H}_{\text{XXZ}, \phi} = \bigoplus_{j=-L/2}^{L/2} \mathcal{H}^{S^z = j}, \qquad \mathcal{H}^{S^z = j} \simeq W_{|j|, z^2=e^{i\phi}},
\end{equation}
where the $W_{j,z^2}$ are the standard modules of the affine Temperley-Lieb algebra labeled by $j = |S^z|$ and twist parameter $z^2 = e^{i\phi}$.
We can determine $\phi$ by comparing the Hilbert-space decomposition \eqref{eq:HilbTL} of the $SU(3)$-symmetric biquadratic spin-1 chain to that of this twisted spin-$\tfrac12$ XXZ chain. The former can be expanded as
\begin{equation}\label{eq:Hbqdecompose}
    \mathcal{H}_{\text{bq}} \simeq \hat{D}_0 W_{0,q^2} \oplus \hat{D}_1 W_{1,1} \oplus \dots \qquad \text{ with } \hat D_0 = 1, \; \hat D_1 = 7.
\end{equation}
In the $j=0$ sector, the non-contractible loop weight must satisfy $z+z^{-1}=m=3$, so that $z=q^{\pm1}$. The two choices give equivalent modules, $W_{0,q^2}\simeq W_{0,q^{-2}}$. Taking $q=e^\gamma$ and $z=q$, the relation $z^2=e^{i\phi}$ gives the XXZ twist
\begin{equation}\label{eq:twist}
\phi = -2i\gamma = -2i\arccosh(3/2).
\end{equation}
The twist angle \eqref{eq:twist} is purely imaginary, making the twisted XXZ Hamiltonian \eqref{eq:hamXXZ} non-hermitian. Nevertheless, we will find that the low-lying eigenstates, including the ground states, have real energies, as required from their embedding into the hermitian biquadratic spin-1 chain. 

Each eigenvalue in the $S^z=0$ sector of the XXZ chain at $\Delta=\tfrac32$ with twist \eqref{eq:twist} therefore appears once in the biquadratic chain spectrum. The module $W_{1,1}$ instead corresponds to the $S^z=1$ sector with trivial twist $\phi=0$, whose eigenvalues appear seven times. Because the $SU(3)$ (and Read-Saleur) generators can have momentum $\pi$, the Temperley-Lieb mapping fixes momentum only modulo $\pi$,
\begin{equation}
    k_{\rm bq}=K_{\rm XXZ} \mod{\pi}.
\end{equation}
The seven copies of the XXZ $S^z=1$ sector in the biquadratic chain split with multiplicities $5+2$ between the two momenta differing by $\Delta k_{\rm bq} = \pi$ \citep{Aufgebauer2010QuantumSC}.
As we will show below, a state from the twisted $S^z=0$ sector is exactly degenerate with the state in the untwisted $S^z=1$ sector and completes the $5+3$ octet. The XXZ origin and finite-size momenta of the low-lying states are summarized in Table~\ref{tab:low-energy-mapping} and illustrated in the spectrum in Fig.~\ref{fig:spectrumbiquadratic_new}.

\begin{table}[t]
\centering
\small
\begin{tabular}{@{}ccc@{}}
\toprule
State & XXZ sector $(S^z,\phi)$ & $k_{\rm bq}$ \\
\midrule
$\mathrm{gs}_0$ & $(0,-2i\gamma)$ & $0$ \\
$\mathrm{gs}_1$ & $(0,-2i\gamma)$ & $\pi$ \\
$\mathbf 8_0$
& $7\times(1,0) + 1 \times (0,-2i\gamma)$
& $0\,[5],\ \pi\,[3]$ \\
$\mathbf 8_1$
& $7\times(1,0) + 1 \times (0,-2i\gamma)$
& $\pi\pm2\pi/L\,[5], \pm2\pi/L\,[3]$ \\
$\mathbf 1_0$
& $(0,-2i\gamma)$
& $\pi\pm2\pi/L$ \\
$\mathbf 1_1$
& $(0,-2i\gamma)$
& $\pm4\pi/L$ \\
\bottomrule
\end{tabular}
\caption{
\small XXZ origin and finite-size momenta of the lowest states of the biquadratic chain. Numbers in square brackets denote the number of octet components at the indicated momentum.}
\label{tab:low-energy-mapping}
\end{table}

\section{The low-energy spectrum of the biquadratic chain}
\label{sec:bethe}

Since the biquadratic chain is integrable and possesses an $SU(3)$ symmetry, it is ideally suited for analysis via the Bethe ansatz. 
The mapping to the XXZ chain described above makes our life even easier, as we can make use of already-known results. We first confirm the twofold ground-state degeneracy of the biquadratic chain in the thermodynamic limit \cite{barber1989biquadratic, sorensen1990}. 
Our new result is that the lowest-lying $SU(3)$ singlets and octets become degenerate in this limit. Their degeneracy provides explicit evidence for the expected deconfinement, as they can be interpreted as deconfined domain wall pairs transforming as $3 \otimes \overline 3 = 1 \oplus 8$ under $SU(3)$. We emphasize that this degeneracy is unrelated to the unusual lattice symmetries, as it only occurs in the limit. 

The XXZ Bethe ansatz determines the ground state and excitation energies in the sectors identified in
\eqref{eq:Hbqdecompose}. 
Bethe's ansatz parametrizes states as plane waves over a ``reference" state, which for XXZ we take to be the completely polarized eigenstate with all spins up. The Bethe equations then are coupled equations constraining the momenta $k_j$ of these plane waves. One of many subtleties is that these momenta need not be real. Rather they can form ``string" solutions comprised of a set of complex momenta. Precisely, we define rapidities $\lambda_j$
via 
\begin{align}
\exp{i k_j} = \frac{\sin{\left( \lambda_j + i\gamma/2 \right)}}{\sin{\left( \lambda_j - i\gamma/2 \right)}}, 
\end{align}
where we use the standard parameterization
\begin{equation*}
    \Delta = \cosh{\gamma}.
\end{equation*}
For the region of interest $\Delta > 1$ and $\gamma$ is real. An $s$-string here then consists of a set of $s$ rapidities with the same real part, namely $\lambda^0_j+i(s-1-2r)\tfrac{\gamma}2$ with $r=0,\dots s-1$. 

Our results for low-energy spectra of the chains then can be summarized as follows. 
The two ground states in the twisted $S^z=0$ XXZ sector give the $k=0$ and $k=\pi$ ground states of the biquadratic chain. Their energy splitting decays exponentially with $L$, although the correlation length is large. Among the excitations, two-hole Bethe states of the untwisted XXZ chain are exactly degenerate with a state containing two holes and a two-string in the twisted XXZ chain. Together, they make up the octet excitation in the biquadratic chain. A second two-hole/two-string state gives the singlet. We show that its energy approaches that of the octet in the thermodynamic limit.

The Bethe equations for the XXZ chain with twist angle $\phi$ \eqref{eq:hamXXZ} and number of sites $L$ read \cite{DEVEGA1985439}:
\begin{equation}\label{eq:bethe}
    \left( \frac{\sin{\left( \lambda_j - i\gamma/2 \right)}}{\sin{\left( \lambda_j + i\gamma/2 \right)}} \right)^L = e^{i\phi} \prod_{k \ne j}^R \frac{\sin{\left( \lambda_j - \lambda_k - i\gamma \right)}}{\sin{\left( \lambda_j - \lambda_k + i\gamma \right)}}, 
\end{equation}
Here $R$ is the number of spin-down excitations relative to the fully polarized state, each labeled by a rapidity $\lambda_j$. The total magnetization is therefore $S^z = L/2 - R$.
Taking the logarithm of Eq.~\eqref{eq:bethe} yields the logarithmic Bethe equations
\begin{equation}\label{eq:logbethe}
\begin{aligned}
    z_L(\lambda_j)
    &=\frac{2\pi I_j}{L},
    \qquad 1\leq j\leq R,\\
    z_L(\lambda_j)
    &=a_1(\lambda_j)
      -\frac{1}{L}\sum_{k\ne j}^R
       a_2(\lambda_j-\lambda_k)
      -\frac{\phi}{L},\\
    a_n(\lambda)
    &=-i\log\left(
    -\frac{\sin(\lambda-in\gamma/2)}
           {\sin(\lambda+in\gamma/2)}
    \right).
\end{aligned}
\end{equation}
The complex rapidities $\lambda_j$ are $\pi$-periodic by definition, so their real parts can be restricted to lie within an interval of length $\pi$ \citep{Virosztek_1984}. With this convention,
\[
a_n'(\lambda)=\frac{2\sinh(n\gamma)}{\cosh(n\gamma)-\cos(2\lambda)}>0
\]
for real \(\lambda\).
The branch cuts of $a_n(\lambda)$ are chosen to run from $in \gamma /2$ to $i \infty$ and from $-i n \gamma/2$ to $-i\infty$ so that $a_n(\lambda)$ is analytic in $|\Im{\lambda}| < n \gamma/2$  \citep{albertini1997free}. For real $\lambda$, $a_n(\lambda)$ is real and monotonically increasing.
The Bethe quantum numbers $I_j$, $1 \leq j \leq R$ must be half-odd integers for even $R$ and integers for odd $R$ (assuming $L$ is even).
The total energy and momentum of a Bethe state are then 
\begin{equation}\label{eq:enmomentum}
\begin{aligned}
    E &= - J \frac{\sinh{\gamma}}{2} \sum_{j=1}^R a_1'(\lambda_j) = -J \sum_{j=1}^R \frac{(\sinh{\gamma})^2}{\cosh{\gamma} - \cos{(2\lambda_j)}}, \\
    K &= \sum_{j=1}^R \big(\pi - a_1(\lambda_j)\big) + \frac{R\phi}{L} =  \pi R - \frac{2\pi}{L} \sum_{j=1}^R I_j.
\end{aligned}
\end{equation}
Consistent with the $\pi$-periodic definition of the rapidities in Eq.~\eqref{eq:bethe}, energy and momentum are also invariant under $\lambda_j \to \lambda_j + \pi$. 

\begin{figure}
    \centering
    \includegraphics{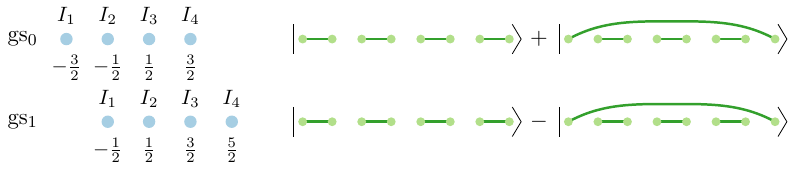}
    \caption{\small Quantum number sets of the two states gs$_0$ and gs$_1$ for an $L = 8$ chain, where gs$_0$ is the symmetric combination of the two dimerized states with zero momentum and gs$_1$ is the anti-symmetric combination with momentum $\pi$.}
    \label{fig:groundstates}
\end{figure}

\paragraph{Ground states}
For the untwisted XXZ chain, \citet{Caux_2008} showed that the quantum numbers $I_j$ of an eigenstate with purely real rapidities $\{ \lambda_j \} \in \mathbb{R}$ can always be chosen from the union of two consecutive intervals $C_0(R) \cup C_1(R)$, where each set is defined by:
\begin{equation}\label{eq:C_i} 
C_i(R)=\left\{
i-\frac{L-R-1}{2},\
i-\frac{L-R-1}{2}+1,\ \ldots,\
i+\frac{L-R-1}{2}
\right\},
\qquad i\in\mathbb Z.
\end{equation}
In the ground state sector, one has $R = L/2$ (i.e., $S^z = 0$), and the two choices 
\begin{align*}
I_j \in C_0(L/2) \; \text{ or } \;
I_j \in C_1(L/2)
\end{align*}
give the two ground states, which we denote gs$_0$ and gs$_1$. In each case, the $L/2$ quantum numbers fill the entire interval $C_i(L/2)$ without gaps. 
Explicitly, 
\begin{equation}
\begin{aligned}\label{eq:A0andA1}
    \text{gs}_0: \; I_j \in C_0(L/2) &= \left\{ -\frac{L}{4} + \frac{1}{2},  -\frac{L}{4} + \frac{3}{2}, \dots, \frac{L}{4} -  \frac{1}{2} \right\}, \\
    \text{gs}_1: \; I_j \in C_1(L/2) &= \left\{ -\frac{L}{4} + \frac{3}{2},  -\frac{L}{4} + \frac{5}{2}, \dots, \frac{L}{4} + \frac{1}{2} \right\}.
\end{aligned}
\end{equation}
Their total momenta differ by $\pi$:
\begin{equation*}
\begin{aligned}
    K_0 &= \pi \frac{L}{2} - \frac{2\pi}{L} \sum_{I_j \in C_0(L/2)} I_j = \pi \frac{L}{2}, \quad
    K_1 = \pi \frac{L}{2} - \pi.
\end{aligned}
\end{equation*}
In a finite chain, these correspond to the symmetric and antisymmetric combinations of the two dimer coverings \cite{affleck1985large}, see Fig.~\ref{fig:groundstates}.

The two ground states are not exactly degenerate at finite size; their energies only become equal in the thermodynamic limit, where the ground state energy per site approaches \cite{Takahashi_1999_chapter4}
\begin{equation}\label{eq:gsenergy_main}
\begin{aligned}
    \lim_{L \to \infty}\frac{E_\text{XXZ}^\text{gs}(J,\Delta)}{L}
    &= -J \sinh{\gamma} \sum_{n = -\infty}^{n = \infty} \frac{1}{1+  e^{2\gamma |n|}},
\end{aligned}
\end{equation} 
independent of the twist $\phi$. 
Using the Temperley--Lieb correspondence at $J=2$ and $\Delta=3/2$, together with $H_{\mathrm{bq}}=H_{\mathrm{TL}}-L$, the thermodynamic ground-state energy density is
\begin{equation}
\begin{aligned}
    \frac{E_{\text{bq}}^\text{gs}}{L}
    &= \frac{E^\text{gs}_\text{XXZ}(J=2,\Delta=3/2)}{L}-1
    = -2.796863434387373\dots .
\end{aligned}
\end{equation}

For finite systems, however, the twist matters: for imaginary $\phi$ as in \eqref{eq:twist}, the rapidities of the ground states with the quantum numbers as above are complex, no longer real, and their energies shift. The difference between the two ground state energies decays exponentially \cite{sorensen1990},
\begin{equation}
    E_{\text{gs}_1} - E_{\text{gs}_0} \sim e^{-L/\xi}, \qquad \xi \approx 42.16.
\end{equation}

\paragraph{Octet excitations}
We now turn to the lowest-lying excited states of the biquadratic chain. As seen from Table 1, for each ground state seven of these occur in the $S^z=1$ sector of the untwisted XXZ chain.
In this sector, $R=L/2-1$, and the lowest-energy excitations of such type are found by putting two holes in the ground-state seas $C_0(R)$ or $C_1(R)$ from Eq.~\eqref{eq:C_i}. These  ``two-spinon" states are thus 
\begin{equation}
\begin{aligned}
    \mathrm{2sp}_0(h_1,h_2):&\qquad
    I_j \in C_0(L/2-1)\setminus\{h_1,h_2\},\\
    \mathrm{2sp}_1(h_1,h_2):&\qquad
    I_j \in C_1(L/2-1)\setminus\{h_1,h_2\}.
\end{aligned}
\end{equation}
%a real-root Bethe state is specified by $R$ occupied quantum numbers drawn from either , as defined in Since each of these sets contains $L-R=L/2+1$ allowed quantum numbers, every such state contains two holes, corresponding to a pair of spin-$\tfrac{1}{2}$ spinons \cite{franchini2017introduction}. We denote a two-spinon state by the positions $h_1,h_2$ of the two holes in the allowed set of Bethe quantum numbers,
The energies of these two-hole states in the thermodynamic limit become \cite{Takahashi_1999_chapter4}
\begin{equation}\label{eq:gap_main}
\begin{aligned}
E_\text{2sp}-E_\text{gs} =E_\text{sp}(\lambda_{h_1}) + E_\text{sp}(\lambda_{h_2}), \quad \text{where}\ \ 
E_\text{sp}(\lambda) = J \sinh{\gamma} \sum_{n=-\infty}^\infty e^{2in\lambda} \frac{e^{-\gamma |n|}}{1 + e^{-2\gamma |n|}}.
\end{aligned}
\end{equation}
The lowest-energy states are thus obtained when the holes lie at the two edges of the Fermi sea, $\lambda_{h_1}=-\pi/2$ and $\lambda_{h_2}=\pi/2$. At $\Delta=3/2$ and $J=2$, their thermodynamic excitation energy is 
\begin{equation}\label{eq:gap}
    \Delta E
    =E_{\text{sp}}(\pi/2)+E_{\text{sp}}(-\pi/2)
    =0.17317888355\ldots\, .
\end{equation}
The 2sp$_0$ and 2sp$_1$ states thus become degenerate in the thermodynamic limit.

Their momentum difference approaches $\pi$, as with the ground states. However, this result only holds in the limit. 
We compare two-spinon states $\mathrm{2sp}_0(h_1,h_2)$ and $\mathrm{2sp}_1(h_1,h_2)$ with holes at the same relative positions in $C_0(R)$ and $C_1(R)$. Each occupied Bethe quantum number then increases by one, $I_j \to I_j + 1$. By Equation~\eqref{eq:enmomentum}, the momentum difference is
\begin{equation}
    K_0-K_1=\frac{2\pi R}{L}
    =\pi-\frac{2\pi}{L}.
\end{equation}
In addition, spatial inversion symmetry of the untwisted XXZ chain implies
%\begin{equation}
    $E(K)=E(-K)$.
%\end{equation}
Thus, for generic $K\neq 0,\pi$, the states at $K$ and $-K$ (whose quantum numbers are related as $I_j \to -I_j$) are distinct but exactly degenerate, whereas no such doubling occurs at the inversion-invariant momenta $K=0,\pi$. 
The doubling is visible in the finite-size spectrum plotted in Fig.~\ref{fig:spectrumbiquadratic_new}, for example in the pair of octet levels at $k=\pi \pm 2\pi/L$.

\begin{figure}
    \centering
    \includegraphics{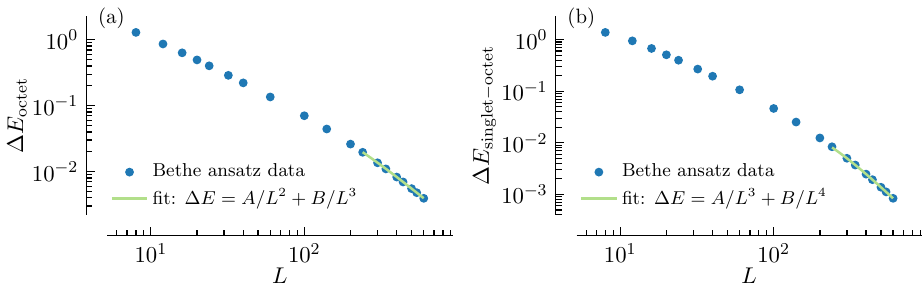}
     \caption{\small
        Finite-size energy splittings obtained from the Bethe ansatz. (a) splitting between the two lowest octet energies.
        (b) splitting between the lowest singlet and the lowest octet at momentum $k=\pi-2\pi/L$.
    }
    \label{fig:octetsplitting}
\end{figure}

We plot in Fig.~\ref{fig:octetsplitting}(a) the finite-size splitting
between the lowest two octets in the biquadratic chain. To obtain these
energies, we solve the logarithmic Bethe equations \eqref{eq:logbethe}
numerically, with one hole at each edge of $C_0(R)$ for the lowest
octet and of $C_1(R)$ for the next.
The energy splitting between these octets scales approximately as $A/L^2+B/L^3$ for $L\gg\xi$. Fitting consecutive triples
$(L,L+20,L+40)$ with $L=300,320,\ldots,560$ gives
$\langle A\rangle\simeq1.61\times10^3$ and
$\langle B\rangle\simeq-1.14\times10^5$, with relative standard deviations $\sigma_A/\langle A\rangle\simeq0.0056$ and
$\sigma_B/|\langle B\rangle|\simeq0.038$.

\paragraph{Singlet and octet excitations with $S^z=0$}
We next consider the lowest-lying $SU(3)$ singlet excited states in the biquadratic chain. As apparent from table 1, these arise from the $S^z=0$ sector of the XXZ chain with twist \eqref{eq:twist}. Octet exciations also occur in this sector, so we must take care to disentangle the two. In the untwisted chain, the lowest excited states in this sector consist of one 2-string and two holes \cite{Virosztek_1984}. 
The quantum numbers of the 1-strings lie in either $C_0(L/2-2)$ or $C_1(L/2-2)$, while $I^2_1$ denotes the quantum number of the 2-string. By the quasi-periodicity of the Bethe-Gaudin-Takahashi equations, it is sufficient to consider $I^2_1=0,1$ \cite{Caux_2008}. In the thermodynamic limit, these correspond to the $SU(2)_q$ singlet and the $S^z=0$ component of the triplet, respectively \cite{Virosztek_1984,Hauer1999xxz,dugave2015form}. In the thermodynamic limit, the energy and momentum contribution of the 2-string is exactly cancelled by a correction to the density of real roots \citep{takahashi_1999}.

The two solutions associated with a given pair of holes remain present after introducing the twist. In the large-$L$ limit, this is reflected in the higher-level Bethe equation, which has two solutions for the 2-string center \cite{dugave2015form,Virosztek_1984}. At the special twist \eqref{eq:twist}, one solution consists of a rapidity $\lambda_{L/2}=-i\infty$ and $L/2-1$ remaining rapidities that obey the Bethe equations of the untwisted $S^z=1$ sector \cite{sorensen1990}. The infinite rapidity contributes no energy, so the two states are exactly degenerate. Under the Temperley-Lieb correspondence, the additional $S^z=0$ state combines with the seven copies of the $S^z=1$ state to form an $SU(3)$ octet, $8=7+1$.

The other low-lying solution has only finite rapidities, and no degenerate $S^z=1$ counterpart. It therefore corresponds to an $SU(3)$ singlet of the biquadratic chain. Its energy approaches that of the octet as $L\to\infty$, producing the asymptotic $8+1=9$ degeneracy at fixed momentum modulo $\pi$. We thus have established deconfinement in the biquadratic chain. 

The finite-size energy difference between octet and singlet states falls off only polynomially with size. As with the 5-plet of the octet states, the
lowest singlet in the biquadratic chain appears at momentum $k=\pi - 2\pi/L$, with its inversion symmetry partner at $k=\pi + 2\pi/L$. 
We compare their energies at finite size by first identifying the Bethe state corresponding to the singlet at small $L$ by matching its Bethe energy with exact diagonalization. The same branch is then continued to larger $L$, using the preceding solution as the initial condition for Newton iteration. We plot the singlet-octet energy difference in Fig.~\ref{fig:octetsplitting}(b) and observe an approximate finite-size scaling as $A/L^3 + B/L^4$ for $L \gg \xi$. To assess the stability of this scaling form, we fit consecutive triples $(L,L+20,L+40)$ with $L=300,320,\ldots,560$. The resulting coefficients have means $\langle A\rangle\simeq2.27\times10^5$ and $\langle B\rangle\simeq-2.76\times10^7$, with relative standard deviations $\sigma_A/\langle A\rangle\simeq0.0066$ and $\sigma_B/|\langle B\rangle|\simeq0.021$.
The singlet–octet splitting therefore vanishes faster than the splitting between the lowest octet levels, but still polynomially.

\section{\texorpdfstring{The $\mathbb{CP}^{m-1}$ nonlinear sigma model at $\theta=\pi$}{The CP(m-1) nonlinear sigma model at theta=pi}}\label{sec:fieldtheory}

In this section we review the $\mathbb{CP}^{m-1}$ nonlinear sigma model at $\theta=\pi$. In the large-$m$ limit, its low-energy states match those of the biquadratic chain, with both exhibiting deconfinement. Namely, we explain how the sigma model has
two vacua and deconfined domain walls in conjugate representations. However, the lattice model is integrable, and we here review the arguments as to why the sigma model is not expected to be.

\subsection{Field theory action, symmetries and anomaly}\label{subsec:CPnsymmetries} 

The $\mathbb{CP}^{m-1}$ nonlinear sigma model can be written in terms of $m$ complex fields $z_i(x,t)$, subject to
\[
    \sum_{i=1}^m |z_i|^2=1,
    \qquad
    z_i \sim e^{i\alpha(x,t)}z_i .
\]
Introducing an auxiliary $U(1)$ gauge field $a_\mu$ for this local phase redundancy, with
\[
    D_\mu z_i=(\partial_\mu-i a_\mu)z_i ,
\]
the Euclidean action is
\begin{align}\label{eq:action}
S = \frac{1}{g^2}\int d^2x \sum_{i=1}^m |D_\mu z_i|^2 + i\theta Q,
\qquad Q= \frac{1}{2\pi}\int d^2x \, \epsilon_{\mu\nu}\partial_\mu a_\nu \in \mathbb Z .
\end{align}
Equivalently, one may use an unnormalized field with
$\sum_i |z_i|^2=R^2$; then the coupling is absorbed into the overall radius of the target-space metric, with $R^2\sim 1/g^2$.

The connection to the biquadratic spin-1 chain comes from the semiclassical
large-representation limit of related antiferromagnetic $SU(m)$ chains. Consider a chain with an $SU(m)$ representation $R$ on even sites and the conjugate representation $\overline R$ on odd sites, with Hamiltonian
\begin{align}\label{eq:Hlargerep}
    H =\sum_{i=1}^{L/2} \sum_{j,k=1}^m (S^j_k)_{2i}\left((\overline S^k_j)_{2i-1}+(\overline S^k_j)_{2i+1}\right).
\end{align}
Here $S^j_k$ are the $SU(m)$ generators in the representation $R$. For $R$ the completely symmetric representation with one row and $n_c$ columns, coherent-state methods in the limit $n_c\to\infty$ give a continuum field valued in
\[
    SU(m)/S(U(1)\times U(m-1)) \simeq \mathbb{CP}^{m-1}.
\]
The topological angle and sigma-model coupling are
\[
    \theta=\pi n_c \mod 2\pi,
    \qquad
    g^2 \sim \frac{1}{n_c}.
\]
The alternating fundamental-antifundamental chain \eqref{eq:hamSUm} 
has $n_c=1$, for which the mapping formally gives $\theta=\pi$. Although this lies outside the controlled large-representation limit, large-$m$ calculations support a qualitative correspondence with the low-energy physics of the $\mathbb{CP}^{m-1}$ model at $\theta=\pi$ \cite{affleck1985large}.

The symmetries of \eqref{eq:action} are slightly subtler than the notation suggests. The apparent $SU(m)$ symmetry acting as $\mathbf{z} \to \mathbf{U} \mathbf{z}$ is not faithful.
The center $\mathbb Z_m\subset SU(m)$ acts on $\mathbf z$ by an overall phase, and is therefore part of the local $U(1)$ gauge redundancy. The faithful global continuous symmetry is consequently
\[
    PSU(m)=SU(m)/\mathbb Z_m .
\]
Since $Q$ is integer-valued, the theory is invariant under $\theta\to\theta+2\pi$.
Charge conjugation acts as
\[
    z_i\to z_i^*, \qquad a_\mu\to -a_\mu ,
\]
and sends $Q\to -Q$, or equivalently $\theta\to-\theta$. The same is true for
orientation-reversing spacetime symmetries such as parity or time reversal. These symmetries are therefore present at $\theta=0$ and $\theta=\pi$.

At $\theta=\pi$, the charge-conjugation or time-reversal symmetry has a nontrivial interplay with the $PSU(m)$ symmetry.
For even $m$ there is a mixed anomaly between these symmetries. 
For odd $m$, including the case $m=3$, there is instead a global inconsistency: the
symmetries can be realized anomaly-free at $\theta=0$ and at $\theta=\pi$ separately, but not in a way that is compatible across both points \cite{gaiotto2017theta,ohmori2019sigma}. In either case, a trivially gapped symmetric infrared theory is obstructed. For $m>2$, the expected resolution of the $\mathbb{CP}^{m-1}$ model at $\theta=\pi$ is a gapped phase with two degenerate vacua, corresponding to spontaneous breaking of charge conjugation or time reversal. This was also recently obtained by connecting the sigma model to a deformed $SU(m)_1$ conformal field theory and analyzing the resulting RG flow \cite{Lecheminant:2024lrb}.

\subsection{\texorpdfstring{Large-$m$ expansion and deconfinement at $\theta=\pi$}{Large-m expansion and deconfinement at theta=pi}}\label{subsec:largem}

At large $m$, the low-energy physics of the $\mathbb{CP}^{m-1}$ model can be derived explicitly.
Introducing a Lagrange multiplier field $\alpha$ to impose the constraint $\sum_{i=1}^m |z_i|^2=1$, one expands around the translationally invariant saddle
\[
    \langle a_\mu\rangle=0,
    \qquad
    \langle \alpha\rangle=M^2 ,
\]
where $M$ is the dynamically generated mass scale. Thus the $z_i$ fields acquire a mass $M$ at leading order in the large-$m$ expansion \citep{Polyakov:1987ez}. The leading fluctuations around this saddle generate a kinetic term for the gauge field,
\[
    S_\text{eff}[a]
    \propto
    \frac{m}{M^2}\int d^2x\, F_{\mu\nu}F_{\mu\nu}
    +\cdots 
\]
The resulting gauge field propagator has a pole at $p^2=0$, producing a linear Coulomb potential with string tension of order $M^2/m$. At $\theta=0$, this confines the $z_i$ quanta \citep{DAdda1978vbw,Coleman19801n}.

The theta angle has a simple interpretation in this large-$m$ picture. By analogy with the massive Schwinger model, it acts as a background electric field. The vacuum energy is schematically
\[
    E_0(\theta)
    \propto
    \min_{k\in\mathbb Z}(\theta-2\pi k)^2 ,
\]
where $k$ labels the different electric field branches. At $\theta=\pi$, the two branches with $k=0$ and $k=1$ become degenerate, leading to a first-order transition with spontaneous breaking of charge conjugation or time reversal. A $z_i$ particle carries one unit of gauge charge and therefore changes the electric field across its worldline. At $\theta=\pi$ it interpolates between the two degenerate vacua. In this sense, $z_i$ is a domain wall between the two vacua, while $\overline z_i$ is the oppositely oriented anti-domain wall. An alternating pair of $z_i$ and $\overline z_i$ does not experience a long-range Coulomb force, and hence is deconfined. This deconfinement mechanism was first understood in the massive Schwinger model \cite{coleman1976schwinger} and extended to $\mathbb{CP}^{m-1}$ at $\theta=\pi$
with large $m$ \cite{DAdda1978vbw,WITTEN1979285}.
\citet{seiberg1984topology} subsequently demonstrated this behavior at finite $m$ using a strong-coupling expansion of a
two-dimensional Euclidean lattice $\mathbb{CP}^{m-1}$
model, with complex unit vectors on sites and compact
$U(1)$ gauge variables on links. 
One interesting property is that the  pole in the gauge-field propagator persists at $\theta=\pi$ despite deconfinement, as opposed to  
the deconfinement mechanism induced by coupling to massless fermions, either minimally \cite{koberle1987deconfinement} or supersymmetrically \cite{WITTEN1979285}, which removes the Coulomb pole.

This large-$m$ picture agrees with the low-energy physics of the biquadratic spin-1 chain.
The two dimerized lattice ground states play the role of the two vacua at $\theta=\pi$. In both field theory and lattice the lowest-energy excitations are domain walls between them, alternating between   transforming as $3$ and $\overline 3$ under $SU(3)$. A domain-wall pair transforms as $    3\otimes \overline 3 = 1\oplus 8 ,$ giving nine low-lying states above each ground state. As there is no confining energy, widely separated pairs are degenerate, in agreement with the spectrum
derived in Section~\ref{sec:bethe}.

In this identification, one-site lattice translation results in the $\mathbb Z_2$ charge-conjugation symmetry of the field theory, since both exchange the two vacua. In the terminology of Ref.~\cite{cheng2023lieb}, the field-theory charge conjugation is an emanant symmetry of the microscopic lattice translation. The $SU(2)$ member of the family, namely the spin-$\tfrac12$ Heisenberg chain, has a mixed anomaly between translation and $PSU(2)$ spin rotation. For the biquadratic chain and its higher-$m$ $SU(m)$ generalizations, an analogous lattice anomaly or global inconsistency matching that of field theory does not seem to have been established in the literature.

\subsection{Obstructions to quantum integrability}

The $\mathbb{CP}^{m-1}$ nonlinear sigma model is classically integrable. As a symmetric-space sigma model, it admits a Lax representation and infinitely many conserved charges \cite{Eichenherr:1978qa}. 
For $m>2$, however, this integrability is generally not expected to survive quantization. 

A counting argument in \cite{GOLDSCHMIDT1980392} indicates that the classical higher-spin current is not protected at the quantum level. Using the $1/m$ expansion, Ref.~\cite{abdalla1981anomaly} found that a classical nonlocal conserved charge cannot be promoted to a conserved quantum charge; this anomaly was later argued to persist to all orders in $1/m$ \citep{Gomes:1982qh}. 
The $1/m$ analysis also indicates particle production and non-factorized scattering \cite{DAdda1978vbw}, in tension with relativistic quantum integrability. However, these works do not analyze the special point $\theta=\pi$ separately.
More recently, the quantum integrability index of Ref.~\cite{Komatsu:2019hgc} gave no positive evidence for quantum integrability in $\mathbb{CP}^{m-1}$ for $m>2$ at any $\theta$, although again it does not prove non-integrability.

Factorized scattering also seems obstructed. The known $SU(m)$-invariant factorized scattering matrices \citep{Berg:1977dp} do not provide an obvious candidate for the domain wall and anti-domain wall scattering in the $\mathbb{CP}^{m-1}$ model at $\theta=\pi$. The symmetries require that the kink-antikink scattering matrices are of Temperley-Lieb form with parameter $m$. However, in a relativistic theory like $\mathbb{CP}^{m-1}$, this $S$ matrix should obey crossing symmetry, and imposing this constraint seems to give unphysical rapidity periodicities \cite{Fendley:2002ww}. 
Moreover, the large-$m$ scattering amplitudes derived in Appendix~\ref{sec:appendix} do not appear to match the integrable Temperley--Lieb form of the $S$ matrix.

Although none of these results conclusively rules out integrability, especially at $\theta=\pi$, the existing evidence gives little support for the model being quantum integrable.
If the $\mathbb{CP}^{m-1}$ nonlinear sigma model for $m > 2$ at $\theta=\pi$ is indeed not integrable, then the integrability of the $SU(m)$ spin chain \eqref{eq:hamSUm} is not preserved by the large-representation continuum limit. 
This property is already plausible before taking the continuum limit. The large-representation spin chains \eqref{eq:Hlargerep} take $n_c>1$, where the local representations are no longer fundamental. The nearest-neighbor interaction then involves more than the singlet-projector channel, so the Temperley-Lieb structure is lost, and integrability is likely lost with it. 
The multiplicities in the spectrum of these large-representation spin chains will no longer be the exponentially growing dimensions of the Temperley-Lieb commutant representations, but rather the polynomially growing $SU(m)$ dimensions \cite{Fendley:2002ww}. This indicates a mismatch between the Temperley-Lieb spin chain we started with, and the nonlinear sigma model that is supposed to describe it.
The next section addresses a different scaling limit, designed to preserve the integrability and Temperley-Lieb commutant symmetry.

\section{Integrable perturbation and criticality}\label{sec:perturbation}

Deriving an effective field theory for a spin chain requires identifying a regime with a small energy gap, so that a continuum description becomes valid. 
To obtain a continuum limit while retaining integrability,
we must close the excitation gap without changing the onsite
representations. The simplest way of doing so is to deform the Hamiltonian by adding a higher
conserved charge that preserves the symmetries and
integrability of the chain. Such deformations have been
studied in the XXZ chain \cite{muramoto1999integrable, fabiansnotes}, and through the mapping discussed above can be adapted to the biquadratic spin-1 chain.
Because the perturbation commutes with the Hamiltonian, it changes the spinon dispersion but leaves the eigenstates unchanged. 
At a critical coupling, the domain walls become gapless with a quadratic dispersion, yielding a nonrelativistic continuum limit.
The already-known massive XXZ S-matrix therefore provides a candidate S-matrix for the field theory near criticality.

\subsection{Perturbing by a higher conserved charge}

Integrable models by definition possess an infinite hierarchy of higher charges $H^{(k)}$ satisfying $[H^{(j)},H^{(k)}]=0$. In a Yang-Baxter integrable model like these, these charges can be written as a sum over local operators. Any sum of these charges thus yields a local integrable Hamiltonian.  For the XXZ chain, these  charges of the closed XXZ chain can be written purely in terms of the Temperley-Lieb generators \citep{nienhuis2021all}. Since these expressions are algebraic, they can be evaluated in any presentation of the Temperley-Lieb algebra, including the $SU(m)$ spin chains considered here and in particular the biquadratic chain.

In particular, for periodic boundary conditions, we have
\begin{align}H^{(1)}=-\tfrac12\sum_i e_i\ ,\qquad H^{(2)} \propto \sum_i \big[e_i,e_{i+1}\big]
\end{align}
where we take the normalization to conform with the conventions of \cite{muramoto1999integrable}. It is easy to verify that the two commute by using the Temperley-Lieb algebra \eqref{eq:tl}. However, as $H^{(2)}$ breaks time-reversal and spatial-parity symmetries, perturbing would destroy the physics of interest. We thus perturb by the next charge
\begin{gather}
\label{eq:H3Nienhuis}
    H(\alpha) =H^{(1)}+\alpha H^{(3)},\\
    H^{(3)} =   \tfrac{1}{8}\sum_i
    \left(
    m\left(e_i e_{i+1}+e_{i+1}e_i\right)
    +2\left(
    e_i e_{i+1}e_{i+2}
    +e_{i+2}e_{i+1}e_i
    -e_{i+1}e_i e_{i+2}
    -e_i e_{i+2}e_{i+1}
    \right)
    \right).
\end{gather}
By construction, $H(\alpha)$ remains integrable and preserves all symmetries of the pure chain.
When $H^{(1)}$ is the XXZ Hamiltonian, \citet{muramoto1999integrable} give an explicit expression for $H^{(3)}$ in terms of spin-$\tfrac12$ operators.

Because both $H^{(1)}$ and $H^{(3)}$ can be written in terms of the Temperley-Lieb generators, the same sector-wise mapping of the biquadratic chain to twisted spin-$\tfrac12$ XXZ chains from Section~\ref{subsec:mapping} applies to the perturbed Hamiltonian. We have checked this numerically by comparing the ground-state energies to those of the XXZ chain with twist $\phi = -2i\gamma$ for $|\alpha|<\alpha_{c}$, where the phase-transition points $\pm \alpha_c$ are discussed next.

\begin{comment}

Starting with the untwisted XXZ Hamiltonian \eqref{eq:hamXXZ}, 
\citet{muramoto1999integrable} considered the deformation
\begin{equation}\label{eq:XXZperturbed}
\begin{aligned}
    H(\alpha) =& H^{(1)} + \alpha H^{(3)}, \\
    H^{(1)} =& \sum_{j=1}^L \left[ \spin_j \cdot \spin_{j+1} \right]_\Delta - \frac{\Delta}{4} L\ =\  H_\text{XXZ}(J=1,\Delta, \phi=0), \\
    H^{(3)} =& - \frac{1+\Delta^2}{2} \sum_{j=1}^L \left[ \spin_j \cdot \spin_{j+1} \right]_{\frac{2\Delta}{1+\Delta^2}} + \frac{\Delta}{2}\sum_{j=1}^L \spin_j \cdot \spin_{j+2} \\& 
    + 2\Delta \sum_{j=1}^L \left( \left(\spin_j \cdot \spin_{j+2} \right)\left( \spin_{j+1} \cdot \spin_{j+3} \right) -  \left[ \spin_j \cdot \spin_{j+3} \right]_{\Delta} \left[ \spin_{j+1} \cdot \spin_{j+2} \right]_{ \Delta^{-1} } \right)  + \frac{\Delta}{8} L, 
\end{aligned}
\end{equation}
where  
$S^\alpha_i = \sigma^\alpha_i$, and 
\begin{equation*}
    \left[\spin_i \cdot \spin_j \right]_\Delta = S^x_i S^x_j + S^y_i S^y_j + \Delta S^z_i S^z_j
\end{equation*}
\end{comment}

\subsection{The phase diagram}

\begin{figure}
    \centering
    \includegraphics{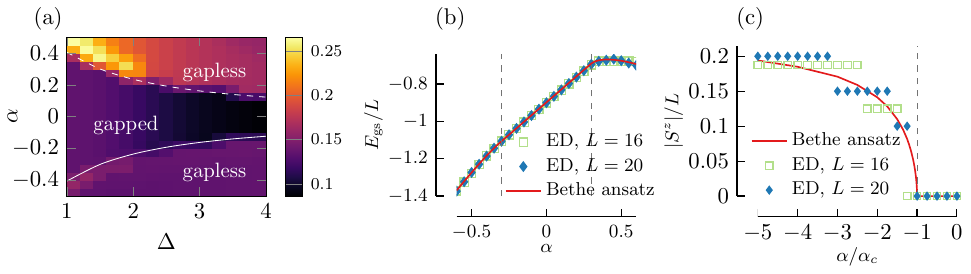}
    \caption{\small (a) Half-cut entanglement entropy phase diagram of the $L = 20$ perturbed XXZ chain with periodic boundary conditions. The transition lines $\pm \alpha_\text{c}$ \eqref{eq:alphac} are plotted in white. (b) Ground state energies and (c) magnetizations of the perturbed XXZ chain at $\Delta = 1.5$. The transitions at $\alpha=\pm \alpha_c$ are marked with dashed gray lines.}
\label{fig:xxzperturbed}
\end{figure}

The phase diagram of the perturbed XXZ chain \eqref{eq:H3Nienhuis} was analyzed in \cite{muramoto1999integrable} using the Bethe ansatz. We focus on the regime $\Delta > 1$ needed to describe the $SU(m)$ spin chains \eqref{eq:hamSUm}. As reviewed in Section~\ref{sec:bethe}, the unperturbed XXZ chain with $\Delta > 1$ is gapped and has two degenerate ground states. This antiferromagnetic gapped phase remains stable up to a critical perturbation strength $|\alpha| < \alpha_c$, given by \cite{muramoto1999integrable}
\begin{equation}\label{eq:alphac}
    \alpha_c = \left( \frac{\gamma}{k K'(k) \sinh{\gamma}} \right)^2.
\end{equation}
Here $K'(k)$ denotes the complete elliptic integral of the first kind with complementary modulus $k' = \sqrt{1-k^2}$, and the modulus $k$ is related to the anisotropy by
\begin{equation*}
\Delta = \cosh{\gamma}, \qquad \gamma = \pi \frac{K'(k)}{K(k)}.
\end{equation*}
Beyond the critical point, $|\alpha| > \alpha_c$, the perturbed chain enters a gapless phase with spontaneous partial magnetization $0 < |\sum_j S^z_j| < L/2$ \citep{muramoto1999integrable}; see also the phase diagram in Fig.~\ref{fig:xxzperturbed}.

The Bethe ansatz also gives the spinon dispersion near $\alpha = \pm \alpha_c$.
The spinon energy of the perturbed chain for $|\alpha| < \alpha_c$ satisfies \cite{muramoto1999integrable}
\begin{align}\label{eq:elambda_pert}
E_{\rm sp}^{(\alpha)}(\lambda)
=
\left[
1+\alpha
\left(
\frac{\sinh\gamma}{2}
\frac{\partial}{\partial\lambda}
\right)^2
\right]
E_{\rm sp}^{(0)}(\lambda).
\end{align}
Here $E_{\mathrm{sp}}^{(0)}$ denotes the spinon dispersion of $H^{(1)}$, equal to half the $J=2$ dispersion in Eq.~\eqref{eq:gap_main}. 

We now derive an explicit expression for the spinon dispersion as a function of momentum, which, to our knowledge, has not been given previously.
Using
\[
    E_{\rm sp}^{(0)}(\lambda)
    =
    -\frac{\sinh\gamma}{2}
    \frac{\partial p(\lambda)}{\partial\lambda},
\]
one can rewrite
\[
    \frac{\sinh\gamma}{2}\frac{\partial}{\partial\lambda}
    =
    -E^{(0)}_{\rm sp}(p)\frac{\partial}{\partial p}.
\]
Therefore
\[
E_{\rm sp}^{(\alpha)}(p)
=
E_{\rm sp}^{(0)}(p)
+
\alpha E_{\rm sp}^{(0)}(p)
\left[
E_{\rm sp}^{(0)}(p)\frac{\partial^2 E_{\rm sp}^{(0)}(p)}{\partial p^2}
+
\left(\frac{\partial E_{\rm sp}^{(0)}(p)}{\partial p}\right)^2
\right].
\]
Inserting
\[
    E_{\rm sp}^{(0)}(p)=
    \frac{K\sinh\gamma}{\pi}
    \sqrt{1-k^2\cos^2 p}
\]
gives
\begin{align}
\label{eq:dispersion}
E_{\rm sp}^{(\alpha)}(p)
=
E_{\rm sp}^{(0)}(p)
\left[
\cos^2p\left(1+\frac{\alpha}{\alpha_c}\right)
+
\sin^2p\left(1-\frac{\alpha}{\alpha_c}\right)
\right].
\end{align}
The lowest physical excitations for a chain with an even number of sites are 2-spinon states, so the physical gap is twice the spinon gap.

\begin{figure}
    \centering
    \includegraphics{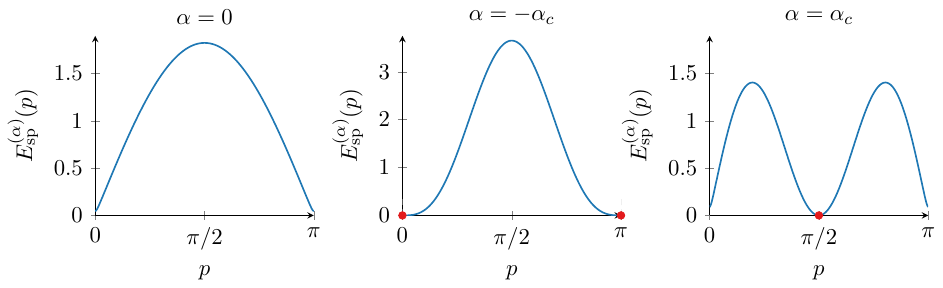}
    \caption{\small Spinon dispersion $E^{(\alpha)}_\mathrm{sp}(p)$ of the perturbed XXZ chain at $\Delta=3/2$ for $\alpha=0$ and at the two critical values $\alpha=\pm \alpha_c$, with $\alpha_c \approx 0.3004$. Gap-closing points are marked with red dots.}
\label{fig:dispersion_perturbation}
\end{figure}

At $\alpha$\,$=$\,$+\alpha_c$, the dispersion \eqref{eq:dispersion} is proportional to
$E_{\rm sp}^{(0)}(p) \cos^2p$, so the gap closes at $p=\pi/2,3\pi/2$. At
$\alpha$\,$=$\,$-\alpha_c$, it is proportional to $E_{\rm sp}^{(0)}(p) \sin^2p$ and the gap closes at $p=0,\pi$. 
Since the dispersion is quadratic near the minimum in both cases, the scaling limit is nonrelativistic rather than Lorentz invariant.
The dispersion relations at the critical points are plotted in Fig.~\ref{fig:dispersion_perturbation}. Since the ground state is unchanged for $|\alpha|<\alpha_c$, its energy
varies linearly with $\alpha$ throughout the gapped phase, as seen in
Fig.~\ref{fig:xxzperturbed}(b).

For the scaling limit we focus on $\alpha=-\alpha_c$, where the gap closes at the original minima of the massive XXZ spinon dispersion. These are the two-spinon states that are the lowest excitations in the unperturbed model. 
The onset of magnetization at $|\alpha|=\alpha_c$ and its behavior for
$|\alpha|>\alpha_c$ can be determined from the thermodynamic Bethe ansatz.
Let $z(\lambda)=\lim_{L\to\infty}z_L(\lambda)$ denote the thermodynamic
counting function introduced in \eqref{eq:logbethe}. Its derivative gives
the total density of Bethe quantum numbers,
\[
\frac{z'(\lambda)}{2\pi}
=
\rho(\lambda)+\rho_h(\lambda),
\]
where $\rho$ and $\rho_h$ are the root and hole densities, respectively. To determine which rapidities are occupied, we introduce the signed TBA
dressed energy $\epsilon_\alpha(\lambda)$, with the convention that $\epsilon_\alpha(\lambda)<0$ for occupied rapidities. For $|\alpha|\leq\alpha_c$, the Bethe sea is completely filled and a spinon is
created by removing an occupied rapidity. Hence its positive excitation energy
is
\[
E_{\rm sp}^{(\alpha)}(\lambda)
=
-\epsilon_\alpha(\lambda).
\]
For $|\alpha|>\alpha_c$, the dressed energy becomes positive on part of the rapidity domain, and the occupied region must instead be determined self-consistently from the zero-temperature TBA equation.
\begin{align}
\epsilon_\alpha(\lambda)
={}&
-\frac{\sinh\gamma}{2}
\left[
1+\alpha
\left(
\frac{\sinh\gamma}{2}
\frac{\partial}{\partial\lambda}
\right)^2
\right]a_1'(\lambda)
\nonumber\\
&-
\int_{-\pi/2}^{\pi/2}\frac{d\mu}{2\pi}\,
a_2'(\lambda-\mu)\epsilon^-_\alpha(\mu),
\qquad
\epsilon^-_\alpha(\lambda)=\min\{\epsilon_\alpha(\lambda),0\}.
\label{eq:dressed-energy-magnetized}
\end{align}
Thus the occupied set is
$D^-=\{\lambda:\epsilon_\alpha(\lambda)<0\}$, and its boundaries are the zeroes of $\epsilon_\alpha(\lambda)$. 
Once $D^-$ has been determined, the root density follows from
\begin{equation}
\rho(\lambda)
=
\mathbf 1_{D^-}(\lambda)
\left[
\frac{a_1'(\lambda)}{2\pi}
-
\int_{-\pi/2}^{\pi/2}\frac{d\mu}{2\pi}\,
a_2'(\lambda-\mu)\rho(\mu)
\right],
\label{eq:root-density-magnetized}
\end{equation}
where $\mathbf 1_{D^-}$ is the indicator function of $D^-$.  The
magnetization is then
\[\frac{|S^z|}{L}
=
\frac12-\int_{D^-}d\lambda\,\rho(\lambda).
\]
We solved these equations on a uniform grid of 4096 rapidities, and plot the result in Fig.~\ref{fig:xxzperturbed}(c). The magnetization has a square root onset as $|S^z|/L\propto\sqrt{|\alpha|/\alpha_c-1}$ when
$\alpha\to-\alpha_c$.

For comparison, consider the gapped antiferromagnetic XXZ chain in a longitudinal magnetic field. The perturbation $h\sum_j\sigma_j^z$ commutes with the zero-field XXZ Hamiltonian and shifts the energies without changing the eigenstates, just as our integrable perturbation. At $h=h_{c_1}$, the
field-driven chain undergoes a Pokrovsky--Talapov transition into a gapless incommensurate phase
\cite{pokrovskytalapov,dzhaparidze1978magnetic,chitra1997critical,cabra1998magnetization}. This transition is characterized by a quadratic dispersion and a square-root onset of the magnetization, just as in our case.
The main difference is that our higher-charge perturbation $H^{(3)}$ preserves the spin-flip symmetry of the XXZ chain. In a magnetic field, the spin-up and spin-down spinons of the XXZ chain are split, so only one species becomes gapless at $h_{c_1}$. Here the splitting is absent, and both species become gapless at $\alpha=-\alpha_c$.

\subsection{Scaling limit near the critical point }\label{subsec:scalinglimit}

The common Temperley-Lieb structure allows the scaling
limits of the perturbed XXZ and biquadratic chains to be
studied together. The same critical points as in the perturbed XXZ chain are expected, with a quadratic low-energy dispersion at $\alpha=\pm \alpha_c$.
For the scaling limit, we again focus on the critical point $\alpha=-\alpha_c$, where the gap closes at $p=0,\pi$, so that the nine lowest-lying states of the unperturbed biquadratic chain become gapless. More generally for the $SU(m)$ Temperley-Lieb chains, all $m$ and $\overline m$ domain walls are expected to become gapless.

We first examine their shared bulk spinon
dispersion and then turn to scattering. The boundary twist can affect finite-size energies, but not the thermodynamic bulk spinon dispersion. 
The critical point $\alpha=-\alpha_c$
is approached from the gapped side, keeping $\gamma$ fixed. The lattice momentum is written $ap$, where $p$ is the
continuum momentum held fixed as the lattice spacing $a\to0$.  At criticality, the dispersion is quadratic, so excitation energies scale as $a^2$.
To retain a finite continuum gap, the distance from criticality must scale at the same rate:
\[
    \delta\alpha=\alpha+\alpha_c=a^2\mu,
    \qquad \mu\geq0 \text{ fixed}.
\]
The continuum limit is obtained by taking $a\to0$ and
$L\to\infty$ at fixed physical length $\ell=La$.
This is analogous in spirit to the massive scaling limit of the antiferromagnetic XXZ chain near the isotropic point \citep{Hauer1999xxz,rutkevich2020scaling}, although the present limit has dynamical exponent $z=2$ instead of $z=1$.

Rescaling
excitation energies by $a^{-2}$ then gives a finite
continuum dispersion. With the above scaling of
$\delta\alpha$, the rescaled spinon dispersion
\eqref{eq:dispersion} takes the form
\begin{align*}
E^{(\alpha)}_{\rm sp}(p)=
\frac{K\sinh\gamma}{\pi a^2}
\sqrt{1-k^2\cos^2(ap)}
\left[
\cos^2(ap)\frac{\delta\alpha}{\alpha_c}
+
\sin^2(ap)\left(2-\frac{\delta\alpha}{\alpha_c}\right)
\right].
\end{align*}
For small $a$,
\[
\sqrt{1-k^2\cos^2(ap)}=k'+O(a^2p^2), 
\]
and
\[
\cos^2(ap)\frac{\delta\alpha}{\alpha_c}
+
\sin^2(ap)\left(2-\frac{\delta\alpha}{\alpha_c}\right)
=
a^2\left(\frac{\mu}{\alpha_c}+2p^2\right)+O(a^4), \qquad \mu :=\frac{\delta\alpha}{a^2}.
\]
Therefore the continuum dispersion takes the nonrelativistic form
\begin{equation}\label{eq:dispersionlimit}
E^{(\alpha)}_{\rm sp}(p)
\longrightarrow
\frac{\mu}{4M\alpha_c}+\frac{p^2}{2M},
\qquad
M=\frac{\pi}{4K k'\sinh\gamma}.
\end{equation}
The kinetic mass $M$ remains finite at criticality,
while the gap vanishes as $\mu\to0$.

\subsection{S-matrix derivation from symmetry constraints}\label{subsec:Smatrix}

Finally, we aim to characterize scattering in the integrable field theory describing the perturbed biquadratic chain near criticality. To this end, $SU(m)$ symmetry, integrability and Galilean invariance are used to constrain the possible S-matrices. 

As a consequence of the deconfinement we demonstrated above, the elementary particles in the field theory transform in the $m$ and $\overline m$ representations of $SU(m)$. The $m$ and $\overline m$ particles correspond to oppositely oriented domain walls between the two ground states and must alternate. We therefore only need to consider $m$--$\overline m$ scattering.
Since $m \otimes \overline m$ decomposes into the singlet and adjoint representations, $SU(m)$ invariance implies that the S-matrix takes the form \cite{Fendley:2002ww}
\begin{equation}
    \check S(p_1,p_2) = f^{(0)}(p_1, p_2) P^{(0)} + f^{(a)}(p_1, p_2) P^{(a)}.
\end{equation}
The projectors onto the singlet and adjoint representations can be written in terms of the Temperley-Lieb generator $e$:
\begin{equation}
    P^{(0)} = \frac{e}{m}, \qquad P^{(a)} = 1-\frac{e}{m}. 
\end{equation}
The Read-Saleur symmetry enlarging $SU(m)$ introduces no additional degeneracies in the two-particle sector. Such larger degeneracies first appear in sectors with four or more particles and are therefore not expected to modify the two-particle S-matrix.

We assume that the field theory is Galilean invariant. The $m$ and $\overline m$ particles have the same mass $M$, as they are exchanged by charge-conjugation symmetry. 
In the scaling limit, $M$ depends on $\gamma$ rather than
the distance from criticality, as shown in
\eqref{eq:dispersionlimit}.
A Galilean boost shifts both momenta by $Mv$, so the S-matrix can depend only on the invariant momentum difference $p_{12}\equiv p_1-p_2$. This constraint along with $SU(m)$ invariance requires the S-matrix be of the form
\begin{equation}
\label{eq:SZF}
    \check S(p_{12}) = Z(p_{12}) \big( 1 + F(p_{12}) e \big).
\end{equation}
Integrability requires that $\check S(p_{12})$ satisfies the Yang-Baxter equation. A simple calculation using the Temperley-Lieb algebra fixes the function $F(p_{12})$ up to a constant $A$ \cite{Fendley:2002ww}:
\begin{equation}\label{eq:SmatrixF}
    F(p_{12}) = - \frac{\sinh{(A p_{12})}}{\sinh{( \gamma + A p_{12})}}, \qquad m = 2 \cosh{\gamma}.
\end{equation}
Unitarity requires $\check S^\dagger(p_{12}) \check S(p_{12}) = 1$, 
a constraint solved by requiring $A$ to be purely imaginary and $Z(p_{12}) Z(-p_{12}) = 1$. 

In a field theory with Lorentz invariance, the S-matrix must satisfy crossing symmetry, allowing it to be fully determined, up to the possibility of including extra poles and the corresponding bound states. As our models are not relativistic even for large $m$, symmetry considerations go no further. The good news is that without crossing symmetry, the difficulties discussed in \cite{Fendley:2002ww} are no longer relevant. Thus having an S-matrix of the form given in \eqref{eq:SZF} and \eqref{eq:SmatrixF} for $m>2$ is consistent in a non-relativistic theory.

A natural candidate for the undetermined scalar prefactor $Z(p_{12})$ comes from the spinon S-matrix of the massive XXZ chain. The higher-charge perturbation commutes with the XXZ Hamiltonian, changing the spinon dispersion while preserving the eigenstates and their scattering phases. The same S-matrix
therefore applies throughout $|\alpha| \leq \alpha_c$.
For a kink and antikink with rapidity difference $\lambda_{12} = \lambda_1 -\lambda_2$, this S-matrix takes the form \cite{davies1993diagonalization, Jimbo:1994qp}
\begin{equation}\label{eq:SmatrixXXZ}
S(\lambda_{12}) = Z(\lambda_{12})
\begin{pmatrix}
1 & 0 & 0 & 0 \\
0 &\frac{\sin\lambda_{12}}{\sin(\lambda_{12}+i\gamma)} & e^{-i\lambda_{12}}\frac{\sin(i\gamma)}{\sin(\lambda_{12}+i\gamma)} &
0 \\ 0 & e^{i\lambda_{12}}\frac{\sin(i\gamma)}{\sin(\lambda_{12}+i\gamma)} &
\frac{\sin\lambda_{12}}{\sin(\lambda_{12}+i\gamma)} &
0 \\
0 & 0 & 0 & 1
\end{pmatrix}.
\end{equation}
To compare this S-matrix with the symmetry-based ansatz, it is useful to expose the Temperley-Lieb structure,
\[
S(\lambda_{12}) = \check S(\lambda_{12}) P, \qquad \check S(\lambda_{12}) = Z(\lambda_{12}) \left(\mathbb I-
\frac{\sin\lambda_{12}}{\sin(\lambda_{12}+i\gamma)}\,e
\right),
\]
where $e$ denotes the TL generator of the XXZ chain and $P$ a permutation matrix.
The prefactor $Z(\lambda_{12})$ can be written in terms of q-Gamma functions \cite{davies1993diagonalization, dugave2015form}
\begin{equation}\label{eq:prefactor2}
Z(\lambda_{12}) =
-e^{-i\lambda_{12}}
\frac{
\Gamma_{q^{-4}}\left(\frac12+\frac{i\lambda_{12}}{2\gamma}\right)
\Gamma_{q^{-4}}\left(1-\frac{i\lambda_{12}}{2\gamma}\right)}
{
\Gamma_{q^{-4}}\left(\frac12-\frac{i\lambda_{12}}{2\gamma}\right)
\Gamma_{q^{-4}}\left(1+\frac{i\lambda_{12}}{2\gamma}\right)},
\qquad q=e^{\gamma}.
\end{equation}
The overall sign is chosen so that $S(0)=-P$, which matches the convention of Ref.~\cite{davies1993diagonalization}.

The matrix part of the XXZ S-matrix agrees with the symmetry-based ansatz derived above upon identifying $Ap_{12}=-i\lambda_{12}$ and replacing the XXZ Temperley-Lieb generator by $e=mP^{(0)}$. Here $p_{12}$ is the continuum momentum difference and
$\lambda_{12}$ the XXZ lattice rapidity difference.
For the biquadratic chain, $e=3P^{(0)}$ is the $9\times9$ generator in Eq.~\eqref{eq:Pbq}.
Therefore, we propose to treat the Temperley-Lieb S-matrix with the XXZ prefactor and the above identifications as the S-matrix of a  nonrelativistic field theory.

The prefactor satisfies a crossing-type functional relation that sends $\lambda \to -\lambda-i\gamma$ (instead of the relativistic $\lambda \to i \pi - \lambda$). The XXZ S-matrix also satisfies the periodicity relation $\check S(\lambda+\pi) = -\check S(\lambda)$.  
Related $q$-deformed S-matrices with real $q$ have been studied in the AdS/CFT literature
\cite{Arutyunov:2013ega, Arutyunov:2014wdg, Appadu:2018ioy}, although those S-matrices generally are not Galilean invariant.
For comparison with the sigma model,
Appendix~\ref{sec:appendix} derives the large-$m$ scattering amplitudes of the $\mathbb{CP}^{m-1}$ model at $\theta=\pi$.
The resulting amplitudes differ from the candidate S-matrix proposed here.

\section{Conclusions}

We studied the antiferromagnetic biquadratic spin-1 chain as the $m=3$ member of a larger family of integrable $SU(m)$-symmetric Temperley-Lieb chains. Using the sector-by-sector mapping to twisted spin-$\tfrac12$ XXZ chains and the Bethe ansatz, we showed the domain-wall excitations are deconfined. 
Staggering the couplings selects one of the two ground states and confines the domain walls. These results agree with the qualitative large-$m$ picture of the $\mathbb{CP}^{m-1}$ nonlinear sigma model at $\theta=\pi$. 
However, we found that the correspondence stops with these basic properties: the sigma model does not seem to provide an effective field-theory description of the chain. As we explained, the $\mathbb{CP}^{m-1}$ models with $m>2$ are believed to be non-integrable, and so cannot arise in the limit of an integrable chain. %One reason why is that the large-representation limit leading to the sigma-model description loses the Temperley-Lieb structure and likely does not preserve integrability. Indeed, 
Our results lend further support to this belief, although a definitive proof is still lacking.

To probe further, we considered a different continuum limit, obtained by perturbing the chain with a higher conserved charge. Because this charge can be expressed entirely in terms of Temperley-Lieb generators, the perturbation preserves both integrability and the sector-wise correspondence with the XXZ chain. At a critical perturbation strength, the spinon gap closes quadratically, giving a transition with dynamical exponent $z=2$ and a square-root onset of the magnetization. This resembles the Pokrovsky-Talapov transition of the XXZ chain in a magnetic field. Unlike a magnetic field, our perturbation preserves spin-flip symmetry, so all low-lying excitations become critical.

In the region of the transition point, we find a massive nonrelativistic continuum theory. It therefore cannot be described by the relativistic $\mathbb{CP}^{m-1}$ nonlinear sigma model. Since the perturbation changes the spinon dispersion but not the eigenstates, lattice scattering remains governed by the antiferromagnetic XXZ S-matrix. 
For the field theory, $SU(m)$ symmetry, Galilean invariance
and the Yang-Baxter equation give the same Temperley-Lieb
S-matrix structure but leave the scalar prefactor undetermined.
Choosing the XXZ prefactor then gives a candidate for the field theory S-matrix.

Our results illustrate how distinct continuum limits of the same family of spin chains can lead to field theories with fundamentally different kinematics and integrability properties. 
An open problem is to construct a local Lagrangian for a field theory realizing the non-relativistic scaling limit we have described. To this end, it may be worth considering nonrelativistic deformations of the $\mathbb{CP}^{m-1}$ model at $\theta=\pi$. It is completely unclear though as to how the Read-Saleur symmetry would manifest itself. Perhaps unsurprisingly, resolving one mystery seems to have generated a new one.

\paragraph{Acknowledgements}
We are grateful to Fabian Essler for discussions and for sharing his notes on integrable deformations of XXZ. 
LE acknowledges the use of AI tools (ChatGPT Sol and Astra) for assistance with developing code for the Bethe ansatz numerics in Section~\ref{sec:bethe} and the large-$m$ calculation in Appendix~\ref{sec:appendix}.
PF was supported by the EPSRC under grant EP/X030881/1. LE is supported by the Walter Burke Institute for Theoretical Physics at Caltech. 

\appendix

\section{\texorpdfstring
{Scattering matrix of the $\mathbb{CP}^{m-1}$ nonlinear sigma model
at $\theta=\pi$ for large $m$}
{Scattering matrix of the CP(m-1) nonlinear sigma model
at theta=pi for large m}}\label{sec:appendix}

\citet{Zamolodchikov1978factorized} used the large-$m$ expansion to compute the two-particle S-matrix of the $O(m)$ nonlinear sigma model to order $1/m$.
Here we calculate low-energy kink--antikink scattering in the $\mathbb{CP}^{m-1}$ model at $\theta=\pi$ using the effective two-body Hamiltonian in the large-$m$ approximation.
To our knowledge, this calculation has not previously been carried out. 

Consider a pair $z_i\bar z_j$ with $i\ne j$, which lies in the adjoint channel. In the centre-of-mass frame, the particles have momenta $p,-p$. For $p\ll M$, where $M$ is the particle mass, their motion is described by the effective Hamiltonian \cite{WITTEN1979285}
$$
H_{\mathrm{eff}}
=2M-\frac{1}{2M}\left(\partial_x^2+\partial_y^2\right)
+\frac{e_{\mathrm g}^2}{2}
\left[|x-y|+\frac{\theta}{\pi}(x-y)\right],
\qquad
e_{\mathrm g}^2=\frac{12\pi M^2}{m}.
$$
The $\theta$-dependent term comes from the background electric field \cite{coleman1976schwinger}. At $\theta=\pi$, the potential vanishes when the quark lies to the left of the antiquark ($x<y$) and increases linearly for the opposite ordering. 

At large separation $x\ll y$, we write the wavefunction as
\begin{equation*}
    \psi^{ij}_{x\ll y}(x,y)
    =
    e^{ip_1x+ip_2y}
    +
    S_{\mathrm{adj}}(p_1,p_2)\,
    e^{ip_2x+ip_1y},
\end{equation*}
where $p_1>p_2$ are the incoming momenta and
$S_{\mathrm{adj}}(p_1,p_2)$ is the reflection amplitude.
Solving the Schr\"odinger equation for $H_{\mathrm{eff}}$ at
$\theta=\pi$ gives
\begin{equation}\label{eq:Sadj_largem}
    S_{\mathrm{adj}}(p_1,p_2)
    =
    \frac{
        ik\,\operatorname{Ai}(-k^2)-\operatorname{Ai}'(-k^2)
    }{
        ik\,\operatorname{Ai}(-k^2)+\operatorname{Ai}'(-k^2)
    },
    \qquad
    k=\frac{p_1-p_2}{2(Me_{\mathrm g}^2)^{1/3}}.
\end{equation}
Here $\operatorname{Ai}$ is the Airy function and  $\operatorname{Ai}'$ its derivative.
Similar Airy function solutions were obtained in \citet{Haber:1980uy} for the theory at $\theta=0$.
The reflection amplitude has unit modulus. There is no asymptotic transmitted wave $\psi_{x \gg y}$ because the
potential grows without bound for $x>y$.

For scattering in the singlet channel, the effective Hamiltonian includes an additional contact interaction \cite{coleman1976schwinger},
\begin{equation*}
    H_{\mathrm{singlet}}=
    H_\mathrm{eff}
    +g_s\delta(x-y),
\end{equation*}
where the real constant $g_s$ parametrizes the singlet interaction. For an incoming $z_i\bar z_i$ pair, the wavefunction component with outgoing flavours $z_j\bar z_j$ is
\begin{equation*}
    \psi^{ii\to jj}_{x\ll y}(x,y)
    =
    \delta_{ij}e^{ip_1x+ip_2y}
    +
    \left[
        \delta_{ij}S_{\mathrm{adj}}(p_1,p_2)
        +S_A(p_1,p_2)
    \right]e^{ip_2x+ip_1y}.
\end{equation*}
Here $S_A$ is the amplitude for conversion to a different flavour pair. The scattering amplitude corresponding to the incoming singlet state $\tfrac{1}{\sqrt{m}} \sum_i \ket{i \bar i}$ satisfies
\begin{equation*}
    S_{\mathrm{singlet}}
    =S_{\mathrm{adj}}+mS_A.
\end{equation*}
Solving the Schr\"odinger equation for $H_{\mathrm{singlet}}$ at
$\theta=\pi$ gives
\begin{equation}\label{eq:Ssinglet_largem}
    S_{\mathrm{singlet}}(p_1,p_2)
    =
    \frac{
        (ik+\eta)\operatorname{Ai}(-k^2)
        -\operatorname{Ai}'(-k^2)
    }{
        (ik-\eta)\operatorname{Ai}(-k^2)
        +\operatorname{Ai}'(-k^2)
    },
    \qquad
    \eta=\frac{Mg_s}{(Me_{\mathrm g}^2)^{1/3}},
\end{equation}
with $k$ defined above. 

The total large-$m$ scattering matrix acting on the internal space $m \otimes \bar m$ can then be written as 
\begin{equation*}
\check S(p_{12})=Z(p_{12})\bigl[1+mF(p_{12})P^{(0)}\bigr],\quad
    Z(p_{12})=S_{\mathrm{adj}}(p_{12}),
    \quad
    F(p_{12})=\frac{1}{m}
    \left[\frac{S_{\mathrm{singlet}}(p_{12})}
               {S_{\mathrm{adj}}(p_{12})}-1\right],
\end{equation*}
where $P^{(0)}$ projects onto the singlet representation and $p_{12}=p_1-p_2$.
The same parameterization is used in
Sections~\ref{subsec:scalinglimit} and~\ref{subsec:Smatrix}
for the S-matrix of the biquadratic chain in its field theory limit.

This allows a direct comparison between the large-$m$ result derived here and the proposed Temperley--Lieb S-matrix.
At large $m = 2\cosh\gamma$, we have $\sinh\gamma = \frac{m}{2}\bigl(1 + O(m^{-2})\bigr)$ and $q^{-4} = e^{-4\gamma} \approx m^{-4} \to 0$. Using the $q$-Gamma product representation~\cite[Sec.~5.18(i)]{NIST:DLMF}, the ratio of $\Gamma_{q^{-4}}$ factors in~\eqref{eq:prefactor2} reduces to $1 + O(q^{-2}) = 1 + O(m^{-2})$. Setting $A p_{12} = -i\lambda_{12}$ and expanding~\eqref{eq:prefactor2} and~\eqref{eq:SmatrixF} at large $m$ and small momentum difference $p_{12} \to 0$ then gives
\begin{equation*}
\begin{aligned}
    Z(p_{12}) &= -e^{A p_{12}}\bigl(1 + O(m^{-2})\bigr) = -1 - A p_{12} + O(p_{12}^2, m^{-2}),
    \\
    F(p_{12}) &= \frac{1}{m}\bigl(e^{-2Ap_{12}} - 1\bigr) + O(m^{-3}) = -\frac{2A}{m}\,p_{12} + O(p_{12}^2, m^{-3}).
\end{aligned}
\end{equation*}
Expanding the Airy functions in the scattering amplitudes \eqref{eq:Sadj_largem} and \eqref{eq:Ssinglet_largem} at low momentum $k = \frac{p_{12}}{2(Me_{\mathrm g}^2)^{1/3}} \ll 1$ likewise yields a linear momentum dependence,
\begin{equation*}
\begin{aligned}
    Z_{\text{large-}m}(p_{12}) &= -1 - \frac{i}{r_0(Me_{\mathrm g}^2)^{1/3}}\,p_{12} \; + O(p_{12}^2),
    \\
    F_{\text{large-}m}(p_{12}) &= -\frac{i\eta}{m r_0(r_0+\eta)(Me_{\mathrm g}^2)^{1/3}}\,p_{12} \; + O(p_{12}^2),
\end{aligned}
\end{equation*}
where $r_0 = -\operatorname{Ai}'(0)/\operatorname{Ai}(0) \approx 0.729$. Thus, at low momentum and large $m$, both S-matrices take the same qualitative form: $Z(p_{12}) + 1$ is purely imaginary and linear in $p_{12}$, while $F(p_{12})$ is suppressed by $1/m$ and linear in $p_{12}$. 
But even when the linear order coefficients are matched, the higher order expansions in $p_{12}$ do not agree: at large $m$, the XXZ prefactor $Z(p_{12})$ and the Yang-Baxter function $F(p_{12}) $ are exponentials in $p_{12}$, whereas $Z_{\text{large-}m}(p_{12})$ and $F_{\text{large-}m}(p_{12})$ are rational functions, which leads to different higher-order terms starting at $O(p_{12}^3)$.
This shows directly that the large-$m$ $\mathbb{CP}^{m-1}$ S-matrix at $\theta = \pi$ does not satisfy the Yang-Baxter equation beyond leading order in momentum, and confirms that our integrable  field theory is not the $\mathbb{CP}^{m-1}$ nonlinear sigma model at $\theta=\pi$.

\bibliographystyle{apsrev}
\bibliography{references}

@article{affleck1990exact,
doi = {10.1088/0953-8984/2/2/016},
url = {https://doi.org/10.1088/0953-8984/2/2/016},
year = {1990},
month = {jan},
publisher = {},
volume = {2},
number = {2},
pages = {405},
author = {I Affleck},
title = {Exact results on the dimerisation transition in SU(n) antiferromagnetic chains},
journal = {Journal of Physics: Condensed Matter},
}

@article{Kluemper_1989,
doi = {10.1209/0295-5075/9/8/013},
url = {https://doi.org/10.1209/0295-5075/9/8/013},
year = {1989},
month = {aug},
publisher = {},
volume = {9},
number = {8},
pages = {815},
author = {A. Klümper},
title = {New Results for q-State Vertex Models and the Pure Biquadratic Spin-1 Hamiltonian},
journal = {Europhysics Letters},
}

@article{Affleck:1984ar,
    author = "Affleck, Ian",
    title = "{The Quantum Hall Effect, $\sigma$ Models at $\theta = \pi$ and Quantum Spin Chains}",
    reportNumber = "SACLAY-PhT 84/144",
    doi = "10.1016/0550-3213(85)90353-0",
    journal = "Nucl. Phys. B",
    volume = "257",
    pages = "397--406",
    year = "1985"
}

@article{HALDANE1983464,
title = {Continuum dynamics of the 1-D Heisenberg antiferromagnet: Identification with the O(3) nonlinear sigma model},
journal = {Physics Letters A},
volume = {93},
number = {9},
pages = {464-468},
year = {1983},
issn = {0375-9601},
doi = {https://doi.org/10.1016/0375-9601(83)90631-X},
url = {https://www.sciencedirect.com/science/article/pii/037596018390631X},
author = {F.D.M. Haldane},
}

@article{WITTEN1979285,
title = {Instatons, the quark model, and the 1/N expansion},
journal = {Nuclear Physics B},
volume = {149},
number = {2},
pages = {285-320},
year = {1979},
issn = {0550-3213},
doi = {https://doi.org/10.1016/0550-3213(79)90243-8},
url = {https://www.sciencedirect.com/science/article/pii/0550321379902438},
author = {Edward Witten},
}

@article{read2007enlarged,
  title={Enlarged symmetry algebras of spin chains, loop models, and S-matrices},
  author={Read, Nick and Saleur, Hubert},
  journal={Nuclear Physics B},
  volume={777},
  number={3},
  pages={263--315},
  year={2007},
  publisher={Elsevier}
}

@article{Aufgebauer2010QuantumSC,
  title={Quantum spin chains of Temperley-Lieb type: periodic boundary conditions, spectral multiplicities and finite temperature},
  author={Britta Aufgebauer and Andreas Kl\"umper},
  journal={Journal of Statistical Mechanics: Theory and Experiment},
  year={2010},
  volume={2010},
  pages={P05018}
}

@article{sutherland1975spinchain,
  title = {Model for a multicomponent quantum system},
  author = {Sutherland, Bill},
  journal = {Phys. Rev. B},
  volume = {12},
  issue = {9},
  pages = {3795--3805},
  numpages = {0},
  year = {1975},
  month = {Nov},
  publisher = {American Physical Society},
  doi = {10.1103/PhysRevB.12.3795},
  url = {https://link.aps.org/doi/10.1103/PhysRevB.12.3795}
}

@article{itoi1997spin1,
  title = {Extended massless phase and the Haldane phase in a spin-1 isotropic antiferromagnetic chain},
  author = {Itoi, Chigak and Kato, Masa-Hide},
  journal = {Phys. Rev. B},
  volume = {55},
  issue = {13},
  pages = {8295--8303},
  numpages = {0},
  year = {1997},
  month = {Apr},
  publisher = {American Physical Society},
  doi = {10.1103/PhysRevB.55.8295},
  url = {https://link.aps.org/doi/10.1103/PhysRevB.55.8295}
}

@article{moudgalya2022frag,
  title = {Hilbert Space Fragmentation and Commutant Algebras},
  author = {Moudgalya, Sanjay and Motrunich, Olexei I.},
  journal = {Phys. Rev. X},
  volume = {12},
  issue = {1},
  pages = {011050},
  numpages = {44},
  year = {2022},
  month = {Mar},
  publisher = {American Physical Society},
  doi = {10.1103/PhysRevX.12.011050},
  url = {https://link.aps.org/doi/10.1103/PhysRevX.12.011050}
}

@article{Caux_2008,
	doi = {10.1088/1742-5468/2008/08/p08006},
	url = {https://doi.org/10.1088/1742-5468/2008/08/p08006},
	year = 2008,
	month = {aug},
	publisher = {{IOP} Publishing},
	volume = {2008},
	number = {08},
	pages = {P08006},
	author = {Jean-S{\'{e}}bastien Caux and Jorn Mossel and Isaac P{\'{e}}rez Castillo},
	title = {The two-spinon transverse structure factor of the gapped Heisenberg antiferromagnetic chain},
	journal = {Journal of Statistical Mechanics: Theory and Experiment},
}

@article{Virosztek_1984,
	doi = {10.1088/0305-4470/17/15/020},
	url = {https://doi.org/10.1088/0305-4470/17/15/020},
	year = 1984,
	month = {oct},
	publisher = {{IOP} Publishing},
	volume = {17},
	number = {15},
	pages = {3029--3039},
	author = {A Virosztek and F Woynarovich},
	title = {Degenerated ground states and excited states of the S=1/2anisotropic antiferromagnetic Heisenberg chain in the easy axis region},
	journal = {Journal of Physics A: Mathematical and General},
}

@article{albertini1997free,
  title={The free energy singularity of the asymmetric six-vertex model and the excitations of the asymmetric XXZ chain},
  author={Albertini, Giuseppe and Dahmen, Silvio Renato and Wehefritz, Birgit},
  journal={Nuclear Physics B},
  volume={493},
  number={3},
  pages={541--570},
  year={1997},
  publisher={Elsevier}
}

@article{affleck1985large,
  title = {Large-$n$ Limit of $\mathrm{SU}(n)$ Quantum "Spin" Chains},
  author = {Affleck, Ian},
  journal = {Phys. Rev. Lett.},
  volume = {54},
  issue = {10},
  pages = {966--969},
  numpages = {0},
  year = {1985},
  month = {Mar},
  publisher = {American Physical Society},
  doi = {10.1103/PhysRevLett.54.966},
  url = {https://link.aps.org/doi/10.1103/PhysRevLett.54.966}
}

@article{DEVEGA1985439,
title = {Method for calculating finite size corrections in Bethe ansatz systems: Heisenberg chain and six-vertex model},
journal = {Nuclear Physics B},
volume = {251},
pages = {439-456},
year = {1985},
issn = {0550-3213},
doi = {https://doi.org/10.1016/0550-3213(85)90271-8},
url = {https://www.sciencedirect.com/science/article/pii/0550321385902718},
author = {H.J. {de Vega} and F. Woynarovich},
}

@article{sorensen1990,
  title = {Correlation length of the biquadratic spin-1 chain},
  author = {S\o{}rensen, Erik Schwartz and Young, A. Peter},
  journal = {Phys. Rev. B},
  volume = {42},
  issue = {1},
  pages = {754--759},
  numpages = {0},
  year = {1990},
  month = {Jul},
  publisher = {American Physical Society},
  doi = {10.1103/PhysRevB.42.754},
  url = {https://link.aps.org/doi/10.1103/PhysRevB.42.754}
}

@inbook{takahashi_1999, place={Cambridge}, title={XYZ and eight-vertex models}, DOI={10.1017/CBO9780511524332.006}, booktitle={Thermodynamics of One-Dimensional Solvable Models}, publisher={Cambridge University Press}, author={Takahashi, Minoru}, year={1999}, pages={67–84}}

@inbook{Takahashi_1999_chapter4, place={Cambridge}, title={The XXZ model}, booktitle={Thermodynamics of One-Dimensional Solvable Models}, publisher={Cambridge University Press}, author={Takahashi, Minoru}, year={1999}, pages={46–66}}

@article{kormos2017real,
  title={Real-time confinement following a quantum quench to a non-integrable model},
  author={Kormos, Marton and Collura, Mario and Tak{\'a}cs, Gabor and Calabrese, Pasquale},
  journal={Nature Physics},
  volume={13},
  number={3},
  pages={246--249},
  year={2017},
  publisher={Nature Publishing Group UK London}
}

@article{Martin:1993jka,
    author = "Martin, Paul and Saleur, Hubert",
    title = "{The Blob algebra and the periodic Temperley-Lieb algebra}",
    eprint = "hep-th/9302094",
    archivePrefix = "arXiv",
    reportNumber = "USC-93-009",
    doi = "10.1007/BF00805852",
    journal = "Lett. Math. Phys.",
    volume = "30",
    pages = "189",
    year = "1994"
}

@article{Gainutdinov:2017aom,
    author = "Gainutdinov, Azat M. and Jacobsen, Jesper Lykke and Saleur, Hubert",
    title = "{A fusion for the periodic Temperley-Lieb algebra and its continuum limit}",
    eprint = "1712.07076",
    archivePrefix = "arXiv",
    primaryClass = "hep-th",
    reportNumber = "ZMP-HH/18-1, Hamburger Beitrage zur Mathematik 717, ZMP-HH-18-1, HAMBURGER-BEITRAGE-ZUR-MATHEMATIK-717",
    doi = "10.1007/JHEP11(2018)117",
    journal = "JHEP",
    volume = "11",
    pages = "117",
    year = "2018"
}

@article{coleman1976schwinger,
title = {More about the massive Schwinger model},
journal = {Annals of Physics},
volume = {101},
number = {1},
pages = {239-267},
year = {1976},
issn = {0003-4916},
doi = {https://doi.org/10.1016/0003-4916(76)90280-3},
url = {https://www.sciencedirect.com/science/article/pii/0003491676902803},
author = {Sidney Coleman},
}

@article{chitra1997critical,
  title={{Critical properties of gapped spin-chains and ladders in a magnetic field}},
  author={Chitra, R and Giamarchi, T},
  journal={Physical Review B},
  volume={55},
  number={9},
  pages={5816},
  year={1997},
  publisher={APS}
}

@article{cabra1998magnetization,
  title={Magnetization plateaux in N-leg spin ladders},
  author={Cabra, Daniel Carlos and Honecker, Andreas and Pujol, Pierre},
  journal={Physical Review B},
  volume={58},
  number={10},
  pages={6241},
  year={1998},
  publisher={APS}
}

@article{dzhaparidze1978magnetic,
  title={{Magnetic-field phase transition in a one-dimensional system of electrons with attraction}},
  author={Dzhaparidze, GI and Nersesyan, AA},
  journal={Soviet Journal of Experimental and Theoretical Physics Letters},
  volume={27},
  pages={334},
  year={1978}
}

@article{Lecheminant:2024lrb,
    author = "Lecheminant, Philippe and Tanizaki, Yuya and Totsuka, Keisuke",
    title = "{Infrared properties of two-dimensional SU(N)/H nonlinear $\sigma$ models at nonzero $\theta$ angles}",
    eprint = "2412.17493",
    archivePrefix = "arXiv",
    primaryClass = "cond-mat.str-el",
    reportNumber = "YITP-24-171",
    doi = "10.21468/SciPostPhys.18.6.183",
    journal = "SciPost Phys.",
    volume = "18",
    number = "6",
    pages = "183",
    year = "2025"
}

@article{gaiotto2017theta,
  title={Theta, time reversal and temperature},
  author={Gaiotto, Davide and Kapustin, Anton and Komargodski, Zohar and Seiberg, Nathan},
  journal={Journal of High Energy Physics},
  volume={2017},
  number={5},
  pages={1--50},
  year={2017},
  publisher={Springer}
}

@article{ohmori2019sigma,
  title={Sigma models on flags},
  author={Ohmori, Kantaro and Seiberg, Nathan and Shao, Shu-Heng},
  journal={SciPost Physics},
  volume={6},
  number={2},
  pages={017},
  year={2019}
}

@article{Komatsu:2019hgc,
    author = "Komatsu, Shota and Mahajan, Raghu and Shao, Shu-Heng",
    title = "{An Index for Quantum Integrability}",
    eprint = "1907.07186",
    archivePrefix = "arXiv",
    primaryClass = "hep-th",
    doi = "10.21468/SciPostPhys.7.5.065",
    journal = "SciPost Phys.",
    volume = "7",
    number = "5",
    pages = "065",
    year = "2019"
}

@book{Polyakov:1987ez,
    author = "Polyakov, A. M.",
    title = "{Gauge Fields and Strings}",
    doi = "10.1201/9780203755082",
    publisher = "Taylor {\&} Francis",
    address = "London",
    year = "1987"
}

@article{Haber:1980uy,
    author = "Haber, H. E. and Hinchliffe, Ian and Rabinovici, E.",
    title = "{The {CP}**(n-1) Model With Unconstrained Variables}",
    reportNumber = "LBL-10519",
    doi = "10.1016/0550-3213(80)90178-9",
    journal = "Nucl. Phys. B",
    volume = "172",
    pages = "458--474",
    year = "1980"
}

@article{Appadu:2018ioy,
    author = "Appadu, Calan and Hollowood, Timothy J. and Price, Dafydd and Thompson, Daniel C.",
    title = "{Quantum Anisotropic Sigma and Lambda Models as Spin Chains}",
    eprint = "1802.06016",
    archivePrefix = "arXiv",
    primaryClass = "hep-th",
    doi = "10.1088/1751-8121/aadc6d",
    journal = "J. Phys. A",
    volume = "51",
    number = "40",
    pages = "405401",
    year = "2018"
}

@incollection{Coleman19801n,
  author = {Coleman, Sidney},
  title = {{$1/N$}},
  booktitle = {Aspects of Symmetry: Selected Erice Lectures},
  chapter = {8},
  pages = {351--402},
  publisher = {Cambridge University Press},
  address = {Cambridge},
  year = {1985},
  doi = {10.1017/CBO9780511565045.009}
}

@article{DAdda1978vbw,
    author = "D'Adda, A. and L{\"u}scher, M. and Di Vecchia, P.",
    title = "{A 1/n Expandable Series of Nonlinear Sigma Models with Instantons}",
    reportNumber = "Print-78-0885 (BOHR INST.)",
    doi = "10.1016/0550-3213(78)90432-7",
    journal = "Nucl. Phys. B",
    volume = "146",
    pages = "63--76",
    year = "1978"
}

@article{GOLDSCHMIDT1980392,
title = {Conservation laws in some two-dimensional models},
journal = {Physics Letters B},
volume = {91},
number = {3},
pages = {392-396},
year = {1980},
issn = {0370-2693},
doi = {https://doi.org/10.1016/0370-2693(80)91004-7},
url = {https://www.sciencedirect.com/science/article/pii/0370269380910047},
author = {Y.Y. Goldschmidt and E. Witten},
}

@article{abdalla1981anomaly,
  title = {Anomaly in the nonlocal quantum charge of the $\mathrm{C}{\mathrm{P}}^{n\ensuremath{-}1}$ model},
  author = {Abdalla, E. and Abdalla, M. C. B. and Gomes, M.},
  journal = {Phys. Rev. D},
  volume = {23},
  issue = {8},
  pages = {1800--1805},
  numpages = {0},
  year = {1981},
  month = {Apr},
  publisher = {American Physical Society},
  doi = {10.1103/PhysRevD.23.1800},
  url = {https://link.aps.org/doi/10.1103/PhysRevD.23.1800}
}

@article{Gomes:1982qh,
    author = "Gomes, M. and Abdalla, E. and Abdalla, M. C. B.",
    title = "{On the Nonlocal Charge of the {CP}$^{(N-1)}$ Model and Its Supersymmetric Extension to All Orders}",
    reportNumber = "IFUSP/P-331",
    doi = "10.1103/PhysRevD.27.825",
    journal = "Phys. Rev. D",
    volume = "27",
    pages = "825",
    year = "1983"
}

@article{pruisken2022super,
  title = {Superuniversality of dimerized $\mathrm{SU}(N+M)$ spin chains},
  author = {Pruisken, A. M. M. and Danu, Bimla and Shankar, R.},
  journal = {Phys. Rev. B},
  volume = {105},
  issue = {15},
  pages = {155111},
  numpages = {9},
  year = {2022},
  month = {Apr},
  publisher = {American Physical Society},
  doi = {10.1103/PhysRevB.105.155111},
  url = {https://link.aps.org/doi/10.1103/PhysRevB.105.155111}
}

@article{readsachdev1989,
  title = {Valence-bond and spin-Peierls ground states of low-dimensional quantum antiferromagnets},
  author = {Read, N. and Sachdev, Subir},
  journal = {Phys. Rev. Lett.},
  volume = {62},
  issue = {14},
  pages = {1694--1697},
  numpages = {0},
  year = {1989},
  month = {Apr},
  publisher = {American Physical Society},
  doi = {10.1103/PhysRevLett.62.1694},
  url = {https://link.aps.org/doi/10.1103/PhysRevLett.62.1694}
}

@article{Fendley:2002ww,
    author = "Fendley, Paul and Read, Nicholas",
    title = "{Exact S matrices for supersymmetric sigma models and the Potts model}",
    eprint = "hep-th/0207176",
    archivePrefix = "arXiv",
    doi = "10.1088/0305-4470/35/50/301",
    journal = "J. Phys. A",
    volume = "35",
    pages = "0675",
    year = "2002"
}

@article{rutkevich2020scaling,
  title={Scaling in the massive antiferromagnetic X X Z spin-1/2 chain near the isotropic point},
  author={Rutkevich, Sergei B},
  journal={Physical Review E},
  volume={101},
  number={3},
  pages={032115},
  year={2020},
  publisher={APS}
}

@article{cheng2023lieb,
  title={Lieb-Schultz-Mattis, Luttinger, and't Hooft-anomaly matching in lattice systems},
  author={Cheng, Meng and Seiberg, Nathan},
  journal={SciPost Physics},
  volume={15},
  number={2},
  pages={051},
  year={2023}
}

@book{Jimbo:1994qp,
  author = "Jimbo, Michio and Miwa, Tetsuji",
  title = "{Algebraic analysis of solvable lattice models}",
  publisher = "Cambridge University Press",
  year = "1994"
}

@article{davies1993diagonalization,
  title={Diagonalization of the XXZ Hamiltonian by vertex operators},
  author={Davies, Brian and Foda, Omar and Jimbo, Michio and Miwa, Tetsuji and Nakayashiki, Atsushi},
  journal={Communications in Mathematical Physics},
  volume={151},
  number={1},
  pages={89--153},
  year={1993},
  publisher={Springer}
}

@article{dugave2015form,
  title={On form-factor expansions for the XXZ chain in the massive regime},
  author={Dugave, Maxime and G{\"o}hmann, Frank and Kozlowski, Karol K and Suzuki, Junji},
  journal={Journal of Statistical Mechanics: Theory and Experiment},
  volume={2015},
  number={5},
  pages={P05037},
  year={2015},
  publisher={IOP Publishing}
}

@article{nienhuis2021all,
    author = "Nienhuis, Bernard and Huijgen, Onno E.",
    title = "{The local conserved quantities of the closed XXZ chain}",
    eprint = "2104.01851",
    archivePrefix = "arXiv",
    primaryClass = "math-ph",
    doi = "10.1088/1751-8121/ac0961",
    journal = "J. Phys. A",
    volume = "54",
    number = "30",
    pages = "304001",
    year = "2021"
}

@article{affleck2022flag,
  title={Flag manifold sigma models: spin chains and integrable theories},
  author={Affleck, Ian and Bykov, Dmitri and Wamer, Kyle},
  journal={Physics Reports},
  volume={953},
  pages={1--93},
  year={2022},
  publisher={Elsevier}
}

@article{muramoto1999integrable,
  title={Integrable magnetic model of two chains coupled by four-body interactions},
  author={Muramoto, Norihiro and Takahashi, Minoru},
  journal={Journal of the Physical Society of Japan},
  volume={68},
  number={6},
  pages={2098--2104},
  year={1999},
  publisher={The Physical Society of Japan}
}

@article{barber1989biquadratic,
  title = {Spectrum of the biquadratic spin-1 antiferromagnetic chain},
  author = {Barber, Michael N. and Batchelor, Murray T.},
  journal = {Phys. Rev. B},
  volume = {40},
  issue = {7},
  pages = {4621--4626},
  numpages = {0},
  year = {1989},
  month = {Sep},
  publisher = {American Physical Society},
  doi = {10.1103/PhysRevB.40.4621},
  url = {https://link.aps.org/doi/10.1103/PhysRevB.40.4621}
}

@article{solyom1987biquadratic,
  title = {Competing bilinear and biquadratic exchange couplings in spin-1 Heisenberg chains},
  author = {S{\'o}lyom, J.},
  journal = {Phys. Rev. B},
  volume = {36},
  issue = {16},
  pages = {8642--8648},
  numpages = {0},
  year = {1987},
  month = {Dec},
  publisher = {American Physical Society},
  doi = {10.1103/PhysRevB.36.8642},
  url = {https://link.aps.org/doi/10.1103/PhysRevB.36.8642}
}

@article{Temperley1971,
    author = "Temperley, H. N. V. and Lieb, E. H.",
    title = "{Relations between the 'percolation' and 'colouring' problem and other graph-theoretical problems associated with regular planar lattices: some exact results for the 'percolation' problem}",
    doi = "10.1098/rspa.1971.0067",
    journal = "Proc. Roy. Soc. Lond. A",
    volume = "322",
    pages = "251--280",
    year = "1971"
}

@article{ferydouni2026spectrum,
  title={On the Spectrum of Some Temperley-Lieb Spin Chains},
  author={Ferydouni, Robert and Nachtergaele, Bruno},
  journal={arXiv preprint arXiv:2609.38074},
  year={2026}
}

@article{Xian1993dimer,
title = {Exact results of dimerization order parameter in SU (n) antiferromagnetic chains},
journal = {Physics Letters A},
volume = {183},
number = {5},
pages = {437-440},
year = {1993},
issn = {0375-9601},
doi = {https://doi.org/10.1016/0375-9601(93)90602-V},
url = {https://www.sciencedirect.com/science/article/pii/037596019390602V},
author = {Y. Xian},
}

@article{bjornberg2021dimerization,
  title={Dimerization in quantum spin chains with O (n) symmetry},
  author={Bj{\"o}rnberg, Jakob E and M{\"u}hlbacher, Peter and Nachtergaele, Bruno and Ueltschi, Daniel},
  journal={Communications in Mathematical Physics},
  volume={387},
  number={2},
  pages={1151--1189},
  year={2021},
  publisher={Springer}
}

@inproceedings{aizenman2020dimerization,
  title={Dimerization and N{\'e}el Order in Different Quantum Spin Chains Through a Shared Loop Representation: M. Aizenman et al.},
  author={Aizenman, Michael and Duminil-Copin, Hugo and Warzel, Simone},
  booktitle={Annales Henri Poincare},
  volume={21},
  number={8},
  pages={2737--2774},
  year={2020},
  organization={Springer}
}

@article{pasquier1990quantumgroups,
title = {Common structures between finite systems and conformal field theories through quantum groups},
journal = {Nuclear Physics B},
volume = {330},
number = {2},
pages = {523-556},
year = {1990},
issn = {0550-3213},
doi = {https://doi.org/10.1016/0550-3213(90)90122-T},
url = {https://www.sciencedirect.com/science/article/pii/055032139090122T},
author = {V. Pasquier and H. Saleur},
}

@article{levy1991periodic,
  title = {Algebraic structure of translation-invariant spin-1/2 xxz and q-Potts quantum chains},
  author = {Levy, Dan},
  journal = {Phys. Rev. Lett.},
  volume = {67},
  issue = {15},
  pages = {1971--1974},
  numpages = {0},
  year = {1991},
  month = {Oct},
  publisher = {American Physical Society},
  doi = {10.1103/PhysRevLett.67.1971},
  url = {https://link.aps.org/doi/10.1103/PhysRevLett.67.1971}
}

@article{seiberg1984topology,
  title = {Topology in Strong Coupling},
  author = {Seiberg, Nathan},
  journal = {Phys. Rev. Lett.},
  volume = {53},
  issue = {7},
  pages = {637--640},
  numpages = {0},
  year = {1984},
  month = {Aug},
  publisher = {American Physical Society},
  doi = {10.1103/PhysRevLett.53.637},
  url = {https://link.aps.org/doi/10.1103/PhysRevLett.53.637}
}

@article{Arutyunov:2014wdg,
    author = "Arutyunov, Gleb and de Leeuw, Marius and van Tongeren, Stijn J.",
    title = "{The exact spectrum and mirror duality of the $(\text{AdS}_5{\times}S^5)_\eta$ superstring}",
    eprint = "1403.6104",
    archivePrefix = "arXiv",
    primaryClass = "hep-th",
    doi = "10.1007/s11232-015-0243-9",
    journal = "Teor. Mat. Fiz.",
    volume = "182",
    number = "1",
    pages = "28--64",
    year = "2014"
}

@article{Arutyunov:2013ega,
    author = "Arutyunov, Gleb and Borsato, Riccardo and Frolov, Sergey",
    title = "{S-matrix for strings on $\eta$-deformed AdS5 x S5}",
    eprint = "1312.3542",
    archivePrefix = "arXiv",
    primaryClass = "hep-th",
    reportNumber = "ITP-UU-13-31, SPIN-13-23, HU-MATHEMATIK-2013-24, TCD-MATH-13-16",
    doi = "10.1007/JHEP04(2014)002",
    journal = "JHEP",
    volume = "04",
    pages = "002",
    year = "2014"
}

@article{parkinson1987integrability,
doi = {10.1088/0022-3719/20/36/011},
url = {https://doi.org/10.1088/0022-3719/20/36/011},
year = {1987},
month = {dec},
publisher = {},
volume = {20},
number = {36},
pages = {L1029},
author = {J B Parkinson},
title = {On the integrability of the S=1 quantum spin chain with pure biquadratic exchange},
journal = {Journal of Physics C: Solid State Physics},
}

@article{parkinson1988biquadratic,
doi = {10.1088/0022-3719/21/20/014},
url = {https://doi.org/10.1088/0022-3719/21/20/014},
year = {1988},
month = {jul},
publisher = {},
volume = {21},
number = {20},
pages = {3793},
author = {J B Parkinson},
title = {The S=1 quantum spin chain with pure biquadratic exchange},
journal = {Journal of Physics C: Solid State Physics},
}

@article{singh1988spin1,
  title = {Ordering and Criticality in Spin-1 Chains},
  author = {Singh, Rajiv R. P. and Gelfand, Martin P.},
  journal = {Phys. Rev. Lett.},
  volume = {61},
  issue = {18},
  pages = {2133--2136},
  numpages = {0},
  year = {1988},
  month = {Oct},
  publisher = {American Physical Society},
  doi = {10.1103/PhysRevLett.61.2133},
  url = {https://link.aps.org/doi/10.1103/PhysRevLett.61.2133}
}

@article{beard2005cluster,
  title = {Study of $\mathrm{C}\mathrm{P}(N\ensuremath{-}1)$ $\ensuremath{\theta}$-Vacua by Cluster Simulation of $\mathrm{S}\mathrm{U}(N)$ Quantum Spin Ladders},
  author = {Beard, B. B. and Pepe, M. and Riederer, S. and Wiese, U.-J.},
  journal = {Phys. Rev. Lett.},
  volume = {94},
  issue = {1},
  pages = {010603},
  numpages = {4},
  year = {2005},
  month = {Jan},
  publisher = {American Physical Society},
  doi = {10.1103/PhysRevLett.94.010603},
  url = {https://link.aps.org/doi/10.1103/PhysRevLett.94.010603}
}

@article{read1989phasediagram,
title = {Some features of the phase diagram of the square lattice SU(N) antiferromagnet},
journal = {Nuclear Physics B},
volume = {316},
number = {3},
pages = {609-640},
year = {1989},
issn = {0550-3213},
doi = {https://doi.org/10.1016/0550-3213(89)90061-8},
url = {https://www.sciencedirect.com/science/article/pii/0550321389900618},
author = {N. Read and Subir Sachdev},
}

@article{ITensor,
	title={{The ITensor Software Library for Tensor Network Calculations}},
	author={Matthew Fishman and Steven R. White and E. Miles Stoudenmire},
	journal={SciPost Phys. Codebases},
	pages={4},
	year={2022},
	publisher={SciPost},
	doi={10.21468/SciPostPhysCodeb.4},
	url={https://scipost.org/10.21468/SciPostPhysCodeb.4},
}

@article{Zamolodchikov1978factorized,
    author = "Zamolodchikov, Alexander B. and Zamolodchikov, Alexei B.",
    editor = "Khalatnikov, I. M. and Mineev, V. P.",
    title = "{Factorized s Matrices in Two-Dimensions as the Exact Solutions of Certain Relativistic Quantum Field Models}",
    reportNumber = "ITEP-35-1978",
    doi = "10.1016/0003-4916(79)90391-9",
    journal = "Annals Phys.",
    volume = "120",
    pages = "253--291",
    year = "1979"
}

@article{Affleck:1985exact,
    author = "Affleck, Ian",
    title = "{Exact Critical Exponents for Quantum Spin Chains, Nonlinear Sigma Models at Theta = pi and the Quantum Hall Effect}",
    reportNumber = "SACLAY-SPH-T-85-128",
    doi = "10.1016/0550-3213(86)90167-7",
    journal = "Nucl. Phys. B",
    volume = "265",
    pages = "409--447",
    year = "1986"
}

@article{koberle1987deconfinement,
  title = {Deconfinement in the ${\mathrm{CP}}^{\mathrm{N}\mathrm{\ensuremath{-}}1}$ model by massless fermions},
  author = {K{\"o}berle, R. and Kurak, V.},
  journal = {Phys. Rev. Lett.},
  volume = {58},
  issue = {7},
  pages = {627--628},
  numpages = {0},
  year = {1987},
  month = {Feb},
  publisher = {American Physical Society},
  doi = {10.1103/PhysRevLett.58.627},
  url = {https://link.aps.org/doi/10.1103/PhysRevLett.58.627}
}

@misc{NIST:DLMF,
         key = "{\relax DLMF}",
       title = "{\it NIST Digital Library of Mathematical Functions}",
    howpublished = "Release 1.2.8 of 2026-09-15",
         url = "https://dlmf.nist.gov/",
        note = "{F.}~W.~J. Olver, A.~B. {Olde Daalhuis}, D.~W. Lozier, B.~I. Schneider,
                R.~F. Boisvert, C.~W. Clark, B.~R. Miller, B.~V. Saunders,
                H.~S. Cohl, and M.~A. McClain, eds."}

@unpublished{fabiansnotes,
  author       = {Essler, Fabian},
  title        = {Notes on Integrable Spin-1/2 XXZ Chains with 4 and 6-spin Interactions},
  note         = {Unpublished notes},
}

@article{Eichenherr:1978qa,
    author = "Eichenherr, H.",
    title = "{SU(N) Invariant Nonlinear Sigma Models}",
    doi = "10.1016/0550-3213(79)90287-6",
    journal = "Nucl. Phys. B",
    volume = "146",
    pages = "215--223",
    year = "1978",
    note = "[Erratum: Nucl.Phys.B 155, 544 (1979)]"
}

@article{Hauer1999xxz,
    author = "Hauer, T. and Rakos, A and Woynarovich, F",
    title = "{Scaling limit of the one-dimensional XXZ Heisenberg chain with easy axis anisotropy}",
    eprint = "cond-mat/9901184",
    archivePrefix = "arXiv",
    reportNumber = "MIT-CTP-2816",
    doi = "10.1016/S0550-3213(99)00557-X",
    journal = "Nucl. Phys. B",
    volume = "562",
    pages = "497--523",
    year = "1999"
}

@article{Berg:1977dp,
    author = "Berg, B. and Karowski, M. and Weisz, P. and Kurak, V.",
    title = "{Factorized U(n) Symmetric s Matrices in Two-Dimensions}",
    reportNumber = "FUB-HEP 26/77",
    doi = "10.1016/0550-3213(78)90489-3",
    journal = "Nucl. Phys. B",
    volume = "134",
    pages = "125--132",
    year = "1978"
}

@ARTICLE{pokrovskytalapov,
       author = {{Pokrovskii}, V.~L. and {Talapov}, A.~L.},
        title = "{The theory of two-dimensional incommensurate crystals}",
      journal = {Soviet Journal of Experimental and Theoretical Physics},
         year = 1980,
        month = jan,
       volume = {51},
        pages = {134},
       adsurl = {https://ui.adsabs.harvard.edu/abs/1980JETP...51..134P}
}

@article{Affleck1987,
    author = "Affleck, Ian and Haldane, F. D. M.",
    title = "{Critical Theory of Quantum Spin Chains}",
    reportNumber = "PUPT-1052",
    doi = "10.1103/PhysRevB.36.5291",
    journal = "Phys. Rev. B",
    volume = "36",
    pages = "5291--5300",
    year = "1987"
}

@article{Read1990,
    author = "Read, N. and Sachdev, Subir",
    title = "{Spin-Peierls, valence-bond solid, and Neel ground states of low-dimensional quantum antiferromagnets}",
    doi = "10.1103/PhysRevB.42.4568",
    journal = "Phys. Rev. B",
    volume = "42",
    pages = "4568--4589",
    year = "1990"
}

@article{Fendley2000,
    author = "Fendley, Paul",
    title = "{Integrable sigma models with theta = pi}",
    eprint = "cond-mat/0008372",
    archivePrefix = "arXiv",
    doi = "10.1103/PhysRevB.63.104429",
    journal = "Phys. Rev. B",
    volume = "63",
    pages = "104429",
    year = "2001"
}

@article{Blakeney2025,
    author = "Blakeney, Matthew and Corcoran, Luke and de Leeuw, Marius and Pozsgay, Balazs and Vernier, Eric",
    title = "{Temperley-Lieb integrable models and fusion categories}",
    eprint = "2510.19902",
    archivePrefix = "arXiv",
    primaryClass = "cond-mat.str-el",
    doi = "10.1007/JHEP04(2026)165",
    journal = "JHEP",
    volume = "04",
    pages = "165",
    year = "2026"
}

\end{document}